\documentclass[numbers=no]{SciPost}

\hypersetup{
    colorlinks,
    linkcolor={red!50!black},
    citecolor={blue!50!black},
    urlcolor={blue!80!black}
}

\usepackage[bitstream-charter]{mathdesign}
\DeclareSymbolFont{usualmathcal}{OMS}{cmsy}{m}{n}
\DeclareSymbolFontAlphabet{\mathcal}{usualmathcal}

\fancypagestyle{SPstyle}{
\fancyhf{}
\lhead{\colorbox{scipostblue}{\bf \color{white} ~SciPost Physics }}
\rhead{{\bf \color{scipostdeepblue} ~Submission }}

\fancyfoot[C]{\textbf{\thepage}}
}
\usepackage{stackengine}
\usepackage{float}  
\usepackage{tikz-feynman}
\usepackage{xcolor}
\usepackage{tikz}
\usepackage{sectsty}
\usepackage{comment}
\usepackage{scrextend}
\usepackage[normalem]{ulem} 
\usepackage{relsize}
\usetikzlibrary{arrows.meta, automata, positioning, quotes}
\usepackage[normalem]{ulem}
\usepackage{amsmath}
\usepackage{array}
\usepackage{lineno}
\usepackage{makecell}
\usepackage{enumitem}
\usepackage{accents}
\usepackage{ulem}
\usepackage{orcidlink}

\newcommand{\bSigma}{\underline{\Sigma}}   
\newcommand{\bG}{{\underline{G}}}

\newcommand{\bD}{{\underline{D}}}
\newcommand{\bGamma}{{\underline{\Gamma}}}

\newcommand\nver[1]{{\color{black} {#1}}}

\makeatletter
\let\linenumbers\relax

\makeatother

\makeatletter
\DeclareRobustCommand\bigop[2][1]{%
  \mathop{\vphantom{\sum}\mathpalette\bigop@{{#1}{#2}}}\slimits@
}
\newcommand{\bigop@}[2]{\bigop@@#1#2}
\newcommand{\bigop@@}[3]{%
  \vcenter{%
    \sbox\z@{$#1\sum$}%
    \hbox{\resizebox{\ifx#1\displaystyle#2\fi\dimexpr\ht\z@+\dp\z@}{!}{$\m@th#3$}}%
  }%
}

\newcommand{\gaussk}{\DOTSB\bigop[.96]{\mathcal{K}}}

\newcommand\dyckpath[5]{
  \begin{scope}[local bounding box=#4]
    \fill[white]  (#1) rectangle +(#2,#2);
    \fill[red!25!white] (#1) foreach \dir in {#3}{-- ++(\dir*90:1)} |- (#1);
    \path[fill] (#1) foreach \i [count=\j] in {0,...,#5}{ +(\i,0) node[anchor=north]{\j} \ifnum\i>#2 circle (1pt) \fi};
    \draw[help lines] (#1) grid +(#2,#2);
    \draw[line width=2pt] (#1) foreach \dir in {#3}{ -- ++(\dir*90:1)};
  \end{scope}
}

\newcommand\dyckpathh[5]{
  \begin{scope}[local bounding box=#4]
    \fill[white]  (#1) rectangle +(#2,#2);
    
    \fill[red!25!white] (#1) foreach \dir in {#3}{-- ++(\dir*90:1)} |- (#1);
    
    \path[fill] (#1) foreach \i [count=\j] in {0,...,#5}{ +(\i,0) node[anchor=north]{\j} \ifnum\i>#2 circle (1pt) \fi};
    
    \draw[help lines] (#1) grid +(#2,#2);
    
    \draw[line width=2pt] (#1) foreach \dir in {#3}{ -- ++(\dir*90:1)};
    
    \node[star, fill=black, minimum size=6pt, inner sep=0pt] at (#1) {};
    
    \node[star, fill=black, minimum size=6pt, inner sep=0pt] at (#1) ++(#2,0) ++(0,#2) {};
  \end{scope}
}

\tikzfeynmanset{
  fermion1/.style={
    /tikz/postaction={
      /tikz/decoration={
        markings,
        mark=at position 0.5 with {
          \node[
            transform shape,
            xshift=-0.5mm,
            fill,
            dart tail angle=100,
            inner sep=1.3pt,
            draw=none,
            dart
          ] { };
        },
      },
      /tikz/decorate=true,
    },
  },
  fermion2/.style={
    /tikz/postaction={
      /tikz/decoration={
        markings,
        mark=at position 0.65 with {
          \arrow{>[length=6pt, width=5pt]};
        },
      },
      /tikz/decorate=true,
    },
  },
  fermion3/.style={
    /tikz/postaction={
      /tikz/decoration={
        markings,
        mark=at position 0.6 with {
          \arrow{>[length=6pt, width=5pt]};
        },
      },
      /tikz/decorate=true,
    },
  },
}

\definecolor{MatteCyan}{RGB}{95, 158, 160}  
\definecolor{MatteMagenta}{RGB}{139, 0, 139}  
\definecolor{MatteTeal}{RGB}{47, 79, 79}  
\definecolor{MatteOlive}{RGB}{107, 142, 35}  
\definecolor{MatteWine}{RGB}{128, 0, 32}  
\definecolor{MatteGold}{RGB}{184, 134, 11}

\begin{document}
\nolinenumbers
\pagestyle{SPstyle}

\begin{center}{\Large \textbf{\color{scipostdeepblue}{
Polaron formation as the vertex function problem: From Dyck's paths to self-energy Feynman diagrams\\
}}}\end{center}

\begin{center}\textbf{
Tomislav Mi\v{s}ki\'c\orcidlink{0009-0008-6382-5962}\textsuperscript{1,3$\star$},
Juraj Krsnik\orcidlink{0000-0002-4357-2629}\textsuperscript{1}, Stefano Ragni\orcidlink{0009-0003-5603-2968}\textsuperscript{1},
Andrey S. Mishchenko\orcidlink{0000-0002-7626-7567}\textsuperscript{1,2}, and Osor S. Bari\v{s}i\'c\orcidlink{0000-0002-6514-9004}\textsuperscript{1}
}\end{center}

\begin{center}
{\bf 1} Department for Research of Materials under Extreme Conditions, Institute of Physics, HR-10000 Zagreb, Croatia

{\bf 2} RIKEN Center for Emergent Matter Science (CEMS), Wako, Saitama 351-0198, Japan

{\bf 3} Department of Physics, Faculty of Science, University of Zagreb, Bijenička cesta 32, HR-10000 Zagreb, Croatia

$\star$ \href{mailto:tmiskic@ifs.hr}{\small tmiskic@ifs.hr}
\end{center}

\section*{\color{scipostdeepblue}{Abstract}}
\textbf{\boldmath{%
\nver{We present an iterative method for generating the complete set of self-energy Feynman diagrams at arbitrary order for the single-polaron problem with arbitrary linear coupling to the lattice. The approach combines a combinatorial representation of noncrossing diagrams, based on Dyck paths associated with Stieltjes-Rogers polynomials, with the constraints of the Ward-Takahashi identity to systematically incorporate vertex corrections. This construction yields a one-to-one correspondence between terms in the expansion based on Stieltjes-Rogers polynomials and diagrammatic contributions, and provides, through a sequence of simple steps, a closed, algorithmic framework for generating all diagrams of a given order, together with their relative weights. The method enables efficient, unbiased evaluation of diagrammatic series and improves the convergence of diagrammatic Monte Carlo by eliminating the need for stochastic weighting between different topologies. We further outline how the construction can be generalized to finite-density electron systems.}}
}

\vspace{\baselineskip}

\noindent\textcolor{white!90!black}{%
\fbox{\parbox{0.975\linewidth}{%
\textcolor{white!40!black}{\begin{tabular}{lr}%
  \begin{minipage}{0.6\textwidth}%
    {\small Copyright attribution to authors. \newline
    This work is a submission to SciPost Physics. \newline
    License information to appear upon publication. \newline
    Publication information to appear upon publication.}
  \end{minipage} & \begin{minipage}{0.4\textwidth}
    {\small Received Date \newline Accepted Date \newline Published Date}%
  \end{minipage}
\end{tabular}}
}}
}

\linenumbers

\vspace{10pt}
\noindent\rule{\textwidth}{1pt}
\tableofcontents
\noindent\rule{\textwidth}{1pt}
\vspace{10pt}


\section{Introduction}
\label{sec:intro}

\nver{The power of Feynman's diagram technique lies in its ability to rigorously formulate perturbation theory order by order \cite{Mahan}. Individual diagrams are directly associated with specific microscopic quantum processes; identifying the dominant diagrams, therefore, reveals the physical mechanisms governing different regimes of many-body systems. Ideally, the resulting series is summed without approximations and extrapolated to infinite order to obtain exact results.

Purely analytical methods yield exact solutions for only a limited class of problems, and modern approaches increasingly rely on numerical summation of diagrammatic series. One of the most successful techniques is Diagrammatic Monte Carlo (DMC), in which a Metropolis algorithm performs a random walk through different orders and topologies of Feynman diagrams, producing unbiased results for Green's functions and other correlation functions \cite{Prokof2018, Greitemann}. However, stochastic exploration of diagram orders and topologies becomes increasingly inefficient at high orders, as the number of diagrams grows factorially or faster.

{\nver{} The general strategy to improve convergence was first proposed in Ref.~\cite{Rossi}. The central idea is to combine all diagrams of a given order into a single composite object, representing the sum over all topologies with a fixed number of interaction vertices. Such an object is fully specified by the internal imaginary times (or Matsubara frequencies) and momenta (or spatial coordinates) of the vertices. It was shown in Ref.~\cite{Rossi_2017} that the computational cost of this approach is significantly lower than that of treating individual diagrams separately.}

Grouping diagrams by order also mitigates poor convergence associated with sign-alter\-nat\-ing series. For many-fermion systems, the sign problem is typically addressed through determinant expansions and recursive subtraction of disconnected contributions, which are computationally demanding \cite{Simko, Houcke}. In contrast, the perturbation expansion for a single electron coupled to lattice degrees of freedom (the polaron problem) is simpler: fermionic propagators involve one direction of time only, while the phonon propagator remains unrenormalized. Nevertheless, the polaron problem remains well-suited for assessing how diagram-generation schemes can accelerate DMC calculations and, when present, suppress sign fluctuations, providing important insights for further developments.

In this work, we present an algorithm for systematically generating all Feynman diagrams of a given order for the polaron problem with linear coupling to the lattice, outlining a strategy for extending the approach to finite electron densities. The method constructs all possible diagram topologies by combining noncrossing diagrams with vertex-function contributions, yielding composite objects that contain the complete diagrammatic content of each order. This approach substantially reduces the computational cost of high-order calculations and provides a direct means of counting and classifying diagrams at each order. By formulating the method within the simple Holstein model at zero temperature, we are able to expose the universal combinatorial and topological structure of polaron diagrams, which is independent of microscopic details. We further demonstrate, within the Bari\v{s}i\'c-Labb\'e-Friedel and Su-Schrieffer-Heeger (BLF-SSH) model \cite{BLF1, BLF2, BLF3, SSH}, the advantages of grouped diagram sampling in the context of DMC. Indeed, the DMC methods have been successfully applied to a wide range of polaron models \cite{PS98, MPSS, RP, QDST}, including cases with momentum-dependent interaction vertices \cite{tJph, ExPol} that change sign.

The paper is organized as follows. In Section~\ref{sec2}, we express the exact self-energy and its self-consistent Born-Oppenheimer approximation (SCBA) in terms of continued fractions. In Section~\ref{sec3}, we expand the SCBA continued fraction using Stieltjes-Rogers polynomials and provide a graphical representation in terms of Dyck paths, establishing a bijection between the latter and SCBA Feynman diagrams. Section~\ref{sec4} presents a polynomial expansion of the self-energy that includes vertex corrections. In Section~\ref{sec5}, we use the Ward-Takahashi identity to construct the topology of all vertex-function contributions of order $n$ from self-energy diagrams of the same order. Section~\ref{SecSS} introduces a simple iterative algorithm for generating all exact self-energy diagrams from Dyck paths and provides expressions for the number of diagrams. We also provide guidance on extending our method to finite-density systems, including contributions to and from the phonon self-energy. Section~\ref{SecBLF} demonstrates improved convergence of DMC when all diagram topologies at the same order are stochastically sampled simultaneously. Section~\ref{sec7} summarizes our conclusions.}

\section{Polaron problem}
\label{sec2}

\subsection{Exact electron self-energy in the polaron limit}

The exact electron self-energy for the electron-phonon coupled system $\bSigma_k(\omega)$ is shown in Fig.~\ref{f1}. It involves the double solid line representing the exact electron/hole propagator, $\bG_k(\omega)$, and the exact phonon propagator, $\bD_q(\omega)$, represented by the double dashed line. The bare, $g_{k+q,q}$, and the exact vertex, $g_{k+q,q}\bGamma_{k+q,k}(\omega+E,\omega)$ (with $\bGamma$ being the exact vertex function), are represented by the full black circle and the shaded blob, respectively. The self-energy may be expressed in the integral form \cite{Abrikosov},

\begin {equation}
\bSigma_k^{n_e}(\omega)=\frac{i}{N} \sum_q |g_{k+q,q}|^2\int 
\frac{d\omega^\prime}{2\pi} 
\bG_{k+q}(\omega+\omega^\prime)\bD_q(\omega^\prime)\bGamma_{k+q,k}(\omega+\omega^\prime,\omega)\;,\label{SigmaExact}
\end{equation}

\noindent where $n_e$ denotes the density of permanent electrons in the system.  The propagators $\bG_k(\omega)$ and $\bD_q(\omega)$ in Eq.~\eqref{SigmaExact} involve infinite series of reducible diagrams. Furthermore, while there is only one skeleton diagram associated with the exact self-energy $\bSigma_k(\omega)$, the exact vertex involves an infinite series of such diagrams \cite{Fetter}. Thus, the evaluation of Eq.~\eqref{SigmaExact} represents a formidable task, and any simplification or approximation that is justified for some regimes of parameters is a matter of interest \cite{Kabanov,Barisic2006,Loos_2006,Dunn1975,Gumhalter2016}.

\begin{figure}
\[
\vcenter{\hbox{\begin{tikzpicture}
  \begin{feynman}[horizontal = i to o]
    \vertex[dot] (i){};
    \vertex [blob, minimum size = 13pt, right=2.85cm of i] (o){};
    \diagram*{
      (i) --[double,double distance=0.45ex,with arrow=0.5,arrow size=0.165em, edge label' = {$\vec{k}+\vec{q}$, $\omega+\omega^{'}$}] (o); 
      (i) --[double, dashed, double distance=0.45ex, half left, edge label= {$\vec{q}$, $\omega{'}$}] (o);
    };
  \end{feynman}
\end{tikzpicture}}}
\]
\caption{Diagrammatic representation of the exact electron self-energy in terms of the exact electron/hole propagator given by the double solid line, the exact phonon propagator given by the double dashed line, the bare and the exact vertex, given by the full black circle and the shaded blob, respectively.}
\label{f1}
\end{figure}

In the diagrammatic expansion of the ex\-act self-ener\-gy $\bSigma_k^{n_e}(\omega)$, the arguments of all phonon propagators are internal variables, $q$ and $\omega^\prime$, over which the summation/integration should be performed.  The polaron problem corresponds in Eq.~\eqref{SigmaExact} to an electron excitation intermittently added to an otherwise empty band, i.e., $n_e=0$. In this case, within the zero-temperature formalism considered explicitly throughout this paper, the propagator $\bG_k(\omega)$ in Eq.~(\ref{SigmaExact}) represents an electron in the state $k$ propagating forward in time only, with all the poles of $\bG_k(\omega)$, $\bSigma_k(\omega)$, and $\bGamma_{k+q,k}(\omega+\omega^\prime,\omega)$, appearing in the lower complex half-plane of $\omega$. On the other hand, the exact phonon propagator remains unrenormalized since it corresponds to a system with no electrons,

\begin{equation} 
\bD_q(\omega)\rightarrow D_q(\omega)=\frac{1}{\omega-\omega_q+i\eta}-
\frac{1}{\omega+\omega_q-i\eta}\;,\label{barephonon}
\end{equation}

\noindent with $D_q(\omega)$ denoting the bare phonon propagator. With no hole excitations in the system, the integration over internal frequency $\omega^\prime$ in Eq.~(\ref{SigmaExact}) is contributed by the second term in Eq.~(\ref{barephonon}), corresponding to the only pole that is found in the upper complex half-plane of~$\omega^\prime$ in Eq.~(\ref{SigmaExact}), 

\begin{equation}
\bSigma_k(\omega)=\sum_q \frac{|g_{k+q,q}|^2}{N} 
\bG_{k+q}(\omega-\omega_q)
\bGamma_{k+q,k}(\omega-\omega_q,\omega)\;.\label{SigmaPolExact}
\end{equation}

\begin{figure}
\[
\mathlarger{\bSigma_k(\omega)}
~=~
\,\,\mathlarger{\mathlarger{\mathlarger{\sum}}}_{r=0}^{\infty}\,\,\,\,\left(
\vcenter{\hbox{\begin{tikzpicture}
  \begin{feynman}    
    \vertex [dot] (i_2){};
    \vertex [dot, right=0.9cm of i_2] (i_3){};
    \vertex [blob, minimum size = 13pt, right=1.2cm of i_3] (i_4){};
    \vertex [dot, right=1.1cm of i_4] (i_5){};
    \vertex [blob, minimum size = 13pt, right=1.2cm of i_5] (i_6){};
    \vertex [blob, minimum size = 13pt, right=0.9cm of i_6] (i_7){};
    \diagram*{ 
      (i_2) --[fermion2] (i_3);
      (i_3) --[double,double distance=0.35ex,with arrow=0.5,arrow size=0.165em] (i_4);
      (i_4) --[plain, edge label = \(\ldots{}\)] (i_5); 
      (i_5) --[double,double distance=0.35ex,with arrow=0.5,arrow size=0.165em] (i_6);
      (i_6) --[fermion2] (i_7);
    (i_3) --[scalar, half left] (i_4);
    (i_5) --[scalar, half left] (i_6);
    (i_2) --[scalar, half left] (i_7);
    };
  \end{feynman}
  \draw [decorate,decoration={brace,amplitude=4.5pt,mirror,raise=2.15ex}]
  (0.6,0) -- (4.8,0) node[midway,yshift=-2em]{$r$ self-energy insertions $\bSigma_{k+q}(\omega-\omega_q)$};
\end{tikzpicture}}}
\right)
\]
\caption{Diagrammatic content of Eq.~\eqref{cfSigma} for the exact  electron self-energy. Each self-energy insertion $\bSigma_{k+q}(\omega-\omega_q)$ involves an exact electron propagator $\bG_{k+q}(\omega-\omega_q)$, represented by a double line and contributed by all reducible self-energy diagrams. Each self-energy insertion $\bSigma_{k+q}(\omega-\omega_q)$, according to Fig.~\ref{f1} and Eq.~\eqref{GreenDyson}, includes one bare and one exact vertex.}
\label{f2}
\end{figure}

By substituting $\bG_{k+q}(\omega - \omega_q)$ in Eq.~\eqref{SigmaPolExact} using the Dyson equation,

\begin{eqnarray}
\bG_k(\omega) &=& \frac{1}{G^{-1}_k(\omega) - \bSigma_k(\omega)}\;,\label{GreenDyson}
\end{eqnarray}

\noindent the self-energy can formally be rewritten in terms of itself. By recursively repeating this substitution procedure, i.e., each time replacing the exact electron propagator within the self-energy with its own Dyson expansion, the self-energy ultimately takes the form of an infinite continued fraction,

\begin{equation}
\bSigma_k(\omega) = \cfrac{1}{N}\sum_q\frac{ 
|g_{k+q,q}|^2\bGamma_{k+q,k}(\omega-\omega_q,\omega)}
{G_{k+q}^{-1}(\omega-\omega_q)
-\frac{1}{N}\sum_{q'}
\cfrac{|g_{k+q+q',k+q}|^2\bGamma_{k+q+q',k+q}(\omega-\omega_q-\omega_{q'},\omega-\omega_q)}
{G_{k+q+q'}^{-1}(\omega-\omega_q-\omega_{q'}) - \ldots}}\;,\label{cfSigma}
\end{equation}

\noindent where $G_{k}(\omega)$ is the bare electron propagator, 

\begin{equation}
G_{k}(\omega)=\frac{1}{\omega-\varepsilon_k+i\eta}\;,
\label{BareGreen}
\end{equation}

\noindent with $\varepsilon_k$ the electron dispersion. This way of expressing the exact self-energy $\bSigma_k(\omega)$ seems to be absent in the literature. It follows from Eq.~\eqref{cfSigma} that the polaron problem is, in essence, a problem of finding the exact vertex function $\bGamma_{k,k'}(\omega,\omega')$. Furthermore, in Eq.~\eqref{cfSigma}, the summation over momenta $q$ may be generalized to involve a summation over different phonon branches. That is, Eq.~\eqref{cfSigma} is the general equation relating the exact self-energy and the exact vertex function with a single electron characterized by dispersion $\varepsilon_k$, being linearly coupled to lattice phonons \cite{Frohlich_1950,Holstein_1959,Berciu,Bonca_2019,Hahn_2021,Bonca_2022a} (or any bosons \cite{Bergersen,Caruso_2016,Schmitt_Rink_1988,Horsch_1994}) with dispersion~$\omega_q$.

The diagrammatic content of Eq.~\eqref{cfSigma} is illustrated in Fig.~\ref{f2}. One observes that all self-energy diagrams may be represented in the form of noncrossing diagrams, with one of the two vertices in each self-energy insertion being the exact vertex and the other being the bare vertex. Thus, only the vertex function $\bGamma$ involves processes in which the phonon lines may cross, while all other phonon lines remain non-crossing. The rightmost vertex in Fig.~\ref{f2} corresponds to the exact vertex in Fig.~\ref{f1}. The other renormalized vertices in Fig.~\ref{f2} are associated with the renormalization of the exact electron propagator in Eq.~\eqref{SigmaPolExact}, corresponding to an infinite series of reducible self-energy diagrams: $\bG = G + \bSigma\,\bG = G + G\bSigma G + G\bSigma G\bSigma G + \dotsb$. Each self-energy insertion in Fig.~\ref{f2} again contains the exact electron propagator, meaning that, for the polaron problem, the standard diagrammatic expansion that follows from the Wick theorem precisely follows the structure of Fig.~\ref{f2}.

\subsection{Local self-energy} 

 \nver{While the problem of coupled electron-phonon systems demands particular attention to the momentum dependence of vertex corrections \cite{Barisic2007} in order to obtain physically accurate results, several important properties of the diagrammatic expansion are model independent. In particular, the number and topology of diagrams are the same for any electron-phonon model with linear coupling. This allows one to obtain full insight into the topological structure of the diagrammatic expansion without explicitly accounting for momentum dependencies. In practice, it is therefore sufficient to analyze cases in which the exact self-energy is momentum independent (local), $\bSigma_k(\omega)\rightarrow\bSigma(\omega)$.}

In the broader context of polaron physics, there are a few important problems for which the exact self-energy is local. For the Holstein model with local electron-phonon coupling $g_{k,k+q}=g$ and nondispersive optical phonons $\omega_q=\omega_0$, the self-energy becomes local when the vertex corrections are neglected. According to Fig.~\ref{f2}, the remaining diagrams are the noncrossing diagrams, corresponding to the SCBA. Indeed, for the Holstein model, setting $\bGamma=1$ in Eq.~\eqref{cfSigma} gives

\begin{equation}
\Sigma_{SCBA}(\omega) = 
\cfrac{g^2G_{0}(\omega-\omega_0)}
{1-\cfrac{g^{2}G_{0}(\omega-\omega_0)G_{0}(\omega-2\omega_0)}{1-
\cfrac{g^{2}G_{0}(\omega-2\omega_0)G_{0}(\omega-3\omega_0)}{1-\ldots}}}
\;,\label{SSBOA}
\end{equation}

\noindent with $G_0(\omega)=\frac{1}{N}\sum_kG_k(\omega)$, being the bare local propagator, characterized solely by the dispersion of the noninteracting electrons. Thus, with momentum-independent coupling $g$, the SCBA self-energy has no $k$-dependence on any lattice geometry and dimension. 

In the limit of nondispersive electrons (local limit), $\varepsilon_k\rightarrow \varepsilon_0$, the bare electron propagator in Eq.~\eqref{BareGreen} becomes the same as the local propagator $G_0(\omega)$. In this limit, the exact solution of the Holstein polaron problem is given by \cite{Feinberg}

\begin{equation}
\bSigma_0(\omega) = 
\cfrac{g^2G_0(\omega-\omega_0)}
{1-\cfrac{2g^{2}G_0(\omega-\omega_0)G_0(\omega-2\omega_0)}{1-
\cfrac{3g^{2}G_{0}(\omega-2\omega_0)G_{0}(\omega-3\omega_0)}{1-\ldots}}}
\;.\label{SDMFT}
\end{equation}

\noindent Furthermore, $\bSigma_0(\omega)$ is the exact solution of the Holstein impurity problem as well \cite{KrsnikIMP}, characterizing the interaction between the impurity and the electron. The exact self-energy is local in the limit of infinite dimension too. It may be obtained in the context of the Dynamical mean-field theory \cite{Feinberg}, which provides the exact solution by treating the impurity problem self-consistently. In this case, the local propagator in Eq.~\eqref{SDMFT} physically represents the propagation of the electron entering and leaving the same lattice site $m$, $G_0(\omega)\rightarrow G_{m,m'}(\omega)\delta_{m,m'}$. Diagrammatically, $G_{m,m}(\omega)$ involves phonon excitations at other lattice sites as well ($m\neq m'$) \cite{Barisic2007}, and should not be confused with the bare local propagator $G_{0}(\omega)$.

Comparing Eq.~\eqref{SSBOA} and Eq.~\eqref{SDMFT}, it is easy to argue that the SCBA offers a limited improvement over the approximation that considers the leading self-energy contribution only, highlighting the breakdown of the Migdal approximation. Namely, for sufficiently high densities of itinerant charges, characterized by a Fermi energy much larger than the phonon energies, $E_F\gg\omega_q$, the Migdal argument \cite{Migdal} of the unimportance of vertex corrections may be invoked into consideration. However, this argument does not apply to the polaron formation, which is a single-electron problem, $E_F\rightarrow0$. 

\nver{Because Eqs.~\eqref{SSBOA} and \eqref{SDMFT} are formulated in terms of a local self-energy and have a comparatively simple structure, they provide a convenient starting point for analyzing which diagrams appear at successive orders of the general diagrammatic expansion. While Eq.~\eqref{SSBOA} generates only noncrossing self-energy diagrams, Eq.~\eqref{SDMFT} additionally incorporates vertex corrections, thereby encoding the full diagrammatic content at each order. This hierarchy allows us to derive, through a sequence of well-defined steps, an algorithm that constructs an exact, order-by-order classification of all Feynman diagrams by starting from the initial generation of the noncrossing subclass encoded in Eq.~\eqref{SSBOA}.}

\section{Dyck's paths and the SCBA diagrams}
\label{sec3}

\nver{In this section, we expand the SCBA continued fraction in Eq.~\eqref{SSBOA} in terms of Stieltjes-Rogers polynomials \cite{FLAJOLET1980125, Jones_Cfrac, SOKAL_Sfrac, ANGELL_Sfrac_rad, van1993impact}. It is known in the mathematical literature that each term of this expansion admits a graphical representation in terms of Dyck paths. By direct inspection, we find that these Dyck paths are in one-to-one correspondence with the noncrossing self-energy diagrams contributing to the SCBA. This correspondence provides a simple and systematic procedure for generating all topologically inequivalent SCBA (noncrossing) self-energy diagrams and forms the basis for the subsequent analysis.}

\subsection{Stieltjes-Rogers polynomials and Dyck paths}

The general form of the Stieltjes continued fractions, abbreviated as $\mathcal{S}$-fractions, is given by~\cite{FLAJOLET1980125},

\begin{equation}
\mathcal{S}\left(\mathfrak{X}, g\right) = \cfrac{\left.a_0\right|b_1g^2}{1-\cfrac{\left.a_{1}\right|b_{2}g^2}{1-\cfrac{\left.a_{2}\right|b_{3}g^2}{1-\ldots}}}
= \cfrac{\left.a_0\right|b_1g^2}{1+\gaussk_{r = 1}^{+\infty{}}\left(-\left.a_{r}\right|b_{r+1}g^{2}:1\right)}\;.
\label{3}
\end{equation}

\noindent Here, $\gaussk$ is used in a standard way to denote the continued fraction in a more compact form, while $\mathfrak{X}$ stands for any given set of coefficients $a_i$ and $b_i$. The vertical bar $\left.\right|$ means that the denominator is inserted between the factors on the left and right sides during the expansion of the fraction. \nver{The use of these lines is very convenient for the graphical representation of different contributions: more complex contributions are obtained graphically by inserting simpler ones at the correct position to keep the correct order of graphical elements. In particular,} the $\mathcal{S}$-fraction corresponds to an infinite sum \cite{FLAJOLET1980125},

\begin{equation}
\mathcal{S}\left(\mathfrak{X}, z\right) = a_0\cdot{}\sum^{+\infty{}}_{k = 0}R_{2k}\left(\mathfrak{X}\right)g^{2k}\cdot{}b_1g^{2}\;,\label{EqSR}
\end{equation}

\noindent \nver{where $R_{2k}$ are the Stieltjes-Rogers polynomials. In the context of the diagrammatic expansion, $g$ in Eq.~\eqref{EqSR} plays the role of the perturbative (coupling) parameter, with $k$ defining the order of the perturbative expansion, $n = 2k + 2$. The Stieltjes-Rogers polynomials $R_{2k}$ may be expressed in closed form, $R_0\left(\mathfrak{X}\right)=1$, $R_2\left(\mathfrak{X}\right)=a_1b_2$, and}

\begin{equation}
R_{2k>2}\left(\mathfrak{X}\right) = \sum_{\substack{h=0\\ m_0 + \ldots + m_h = k-1}}^{k-1}\,\Biggl(\,\prod_{j = 0}^{h-1}\binom{m_{j}+m_{j+1}-1}{m_j-1}\,\Biggl)\Biggl(\prod_{r = 0}^{h}\left(a_{r+1}b_{r+2}\right)^{m_{j}}\Biggl)\;,
\label{5}
\end{equation}

\noindent \nver{where each valid combination of indices $m_0,\ldots{},m_h$ is defined by the constraint below the sum as given by \cite{FLAJOLET1980125}. Indices $m_0,\ldots{},m_h$ are natural numbers including zero. \par{}It is convenient to analyze the infinite series in Eq.~\eqref{EqSR}, expressed in terms of Stieltjes-Rogers polynomials, through their well-established combinatorial interpretation. As shown in the seminal work of Flajolet~\cite{FLAJOLET1980125}, Stieltjes continued fractions ($\mathcal{S}$-fractions) admit a graphical representation in terms of Dyck paths \cite{BAILEY_Dyck, BRUALDI2008}. This correspondence provides a natural bridge between the algebraic expansion of the continued fraction and the geometric construction. A Dyck path is defined as a sequence of discrete steps in the first quadrant of the x-y plane, built from an up step $a = (1, 1)$ and a down step $b = (1,-1)$. The path starts at $(0,0)$ and ends at $(l,0)$, and is constrained to remain at non-negative height throughout. These conditions imply that each valid path contains an equal number of up and down steps, $n_a = n_b$, so that its total length is $l = n_a + n_b$, with maximal height bounded by $l/2$. We denote by $\mathcal{D}^{(l)}$ the set of all such paths of length $l$.\par{}Within this framework, the indices $m_0,\ldots, m_h$ appearing in Eq.~\eqref{5} acquire a direct combinatorial meaning: each admissible combination uniquely specifies the Dyck path of length $2k$ whose maximal height is $h$. The constraint on the sum of all the components being equal to $k-1$ ensures both the fixed total length and the exclusion of trivial configurations, thereby implementing precisely the Dyck path conditions. Conversely, every Dyck path of fixed length and maximal height determines a unique combination of indices $m_0,\ldots{},m_h$. In our construction, it is useful to attach a height index to each step. This refinement reflects the non-commutative structure of the graphical presentation of the continued-fraction expansion: the contribution associated with a given step depends not only on whether it is the up or down step, but also on the height from which it is taken. In this way, the algebraic structure of the Stieltjes-Rogers polynomials is faithfully encoded in the geometry of the paths.

Truncating the continued fraction at the $h$-th denominator corresponds combinatorially to restricting the Dyck paths to those with maximal height not exceeding $h$. The expansion of the $\mathcal{S}$-fraction may therefore be viewed equivalently as a weighted sum over Dyck paths, with weights determined by the coefficients $a_i$ and $b_i$. For example, if one looks at the first three contributions in Eqs.~(\ref{triv}-\ref{22}), one may see how it corresponds to the expansion of the fraction $a_0\left(1-a_{1}b_{2}g^{2}\right)^{-1}b_1g^2$ up to the first three terms. Analogously, all other terms are obtained by expanding the higher-order fractions.

The Dyck path of length two is given by 
}
\begin{equation}
a_0b_1\equiv{}
	\begin{tikzpicture}[scale=0.7, baseline={([yshift=0.3ex]current bounding box.center)}]

		\fill[red!30] (0,0) -- (1,1) -- (2,0);
		\draw[ultra thick, black] (0,0) -- (1,1) -- (2,0);

        \draw[thin, dashed] node at (-0.33,0.1) {0} (0,0)  -- (2,0) ;
		\draw[thin, dashed] node at (-0.33,1) {1}  (0,1) -- (2,1);
\end{tikzpicture}
    \quad\left[\;
    \begin{tikzpicture}[scale=0.7, baseline={([yshift=-0.4ex]current bounding box.center)}]
  \begin{feynman}
    \vertex(i);
    \vertex  (o) at (-1,0);
    \vertex (f) at (1,0);
    \diagram*{    
      (o) --[scalar, half left] (f);
      (o) --[fermion2] (f);
    };
  \end{feynman} 
\end{tikzpicture}\;\right]
\;.\label{triv}
\end{equation}

\nver{
\noindent In the above equation, we take the up step from height zero, and then the down step from height one, completing the only Dyck path of length two. The dashed lines on the Dyck path diagram are representative measures of the height of the Dyck path on the graph. For later purposes, together with the Dyck paths, we show the corresponding SCBA diagrams in the square brackets. 

Going further, the Dyck path of length four may be represented by a string $a_{0}a_1b_2b_{1}$ which would correspond to the $g^4$ term in the expansion of $a_0\left(1-a_{1}b_{2}g^{2}\right)^{-1}b_1g^2$,
}
\begin{equation}
a_0a_1b_2b_1\equiv{}
\begin{tikzpicture}[scale=0.7, baseline={([yshift=0.3ex]current bounding box.center)}]
		
		\fill[red!30] (0,0) -- (1,1) -- (2,2) -- (3,1) -- (4,0);
		\draw[ultra thick, black] (0,0) -- (1,1) -- (2,2) -- (3,1) -- (4,0);

         \draw[thin, dashed] node at (-0.33,0.1) {0} (0,0)  -- (4,0) ;
		\draw[thin, dashed] node at (-0.33,1) {1}  (0,1) -- (4,1);
		\draw[thin, dashed] node at (-0.33,2) {2}  (0,2) -- (4,2);
\end{tikzpicture}\quad\left[\;
    \begin{tikzpicture}[scale=0.5, baseline={([yshift=-0.4ex]current bounding box.center)}]
  \begin{feynman}
    \vertex(i);
    \vertex  (o) at (-2,0);
    \vertex (f) at (-1,0);
    \vertex (g) at (1,0);
    \vertex (h) at (2,0);
    \diagram*{    
      (o) --[scalar, half left] (h);
      (f) --[scalar, half left] (g);
      (o) --[fermion2] (f);
      (f) --[fermion2] (g);
      (g) --[fermion2] (h);
    };
  \end{feynman} 
\end{tikzpicture}\;\right]
\;.\label{19}
\end{equation}

\noindent \nver{Considering all Dyck paths of length six, we have two contributions given by the $g^6$ terms in the expansion of $a_0\left(1-a_{1}\left(1-a_{2}b_{3}g^{2}\right)^{-1}b_{2}g^{2}\right)^{-1}b_1g^2$,}

\begin{equation}
\begin{aligned}
a_0a_{1}\cdot{}\left(b_{2}a_{1}+a_{2}b_{3}\right)\cdot{}b_{2}b_1&\equiv{}
\begin{tikzpicture}[scale=0.7, baseline={([yshift=0.3ex]current bounding box.center)}]
		
		\fill[red!30] (0,0) -- (1,1) -- (2,2) -- (3,1) -- (4,2) -- (5,1) -- (6,0);
		\draw[ultra thick, black] (0,0) -- (1,1) -- (2,2) -- (3,1) -- (4,2) -- (5,1) -- (6,0);

         \draw[thin, dashed] node at (-0.33,0.1) {0} (0,0)  -- (6,0) ;
		\draw[thin, dashed] node at (-0.33,1) {1}  (0,1) -- (6,1);
		\draw[thin, dashed] node at (-0.33,2) {2}  (0,2) -- (6,2);
        \draw[thin, dashed] node at (-0.33,3) {3}  (0,3) -- (6,3);
\end{tikzpicture}
\quad\left[\;
    \begin{tikzpicture}[scale=0.5, baseline={([yshift=0.3ex]current bounding box.center)}]
  \begin{feynman}
    \vertex (i) at (-3,0);
    \vertex (o) at (-2.,0);
    \vertex (a) at (-0.5,0);
    \vertex (b) at (0.5,0);
    \vertex (f) at (2.,0);
    \vertex (g) at (3,0);
    \diagram*{
      (i) --[scalar, half left] (g);
      (o) --[scalar, half left] (a);
      (b) --[scalar, half left] (f);
      (i) --[fermion2] (o);
      (o) --[fermion2] (a);
      (a) --[fermion2] (b);
      (b) --[fermion2] (f);
      (f) --[fermion2] (g);
    };
  \end{feynman}
\end{tikzpicture}\;\right] \\
& +
\begin{tikzpicture}[scale = 0.7, baseline={([yshift=0.3ex]current bounding box.center)}]
		\fill[red!30] (0,0) -- (1,1) -- (2,2) -- (3,3) -- (4, 2) -- (5, 1) -- (6, 0);
		\draw[ultra thick, black] (0,0) -- (1,1) -- (2,2) -- (3,3) -- (4, 2) -- (5, 1) -- (6, 0);

         \draw[thin, dashed] node at (-0.33,0.1) {0} (0,0)  -- (6,0) ;
		\draw[thin, dashed] node at (-0.33,1) {1}  (0,1) -- (6,1);
		\draw[thin, dashed] node at (-0.33,2) {2}  (0,2) -- (6,2);
        \draw[thin, dashed] node at (-0.33,3) {3}  (0,3) -- (6,3);
\end{tikzpicture}
\quad\left[\;
    \begin{tikzpicture}[scale=0.5, baseline={([yshift=0.3ex]current bounding box.center)}]
  \begin{feynman}
    \vertex (i) at (-3,0);
    \vertex (o) at (-2,0);
    \vertex (a) at (-1,0);
    \vertex (b) at (1,0);
    \vertex (f) at (2,0);
    \vertex (g) at (3,0);
    \diagram*{
      (i) --[scalar, half left] (g);
      (o) --[scalar, half left] (f);
      (a) --[scalar, half left] (b);
      (i) --[fermion2] (o);
      (o) --[fermion2] (a);
      (a) --[fermion2] (b);
      (b) --[fermion2] (f);
      (f) --[fermion2] (g);
    };
  \end{feynman}
\end{tikzpicture}\;\right] 
\;.
\label{22}
\end{aligned}
\end{equation}

\noindent Finally, considering all paths of length eight, one finds five Dyck paths in Eq.~\eqref{26}. For length eight, for the first time, we encounter two distinct Dyck paths with equal contributions, namely the second and third in Eq.~\eqref{26}. Of course, with increasing length of Dyck paths, more paths that have the same contribution appear. 

\nver{The SCBA diagrams, corresponding to Fig.~\ref{f2} with all the vertices being bare, are diagrams without temporal crossing (intersecting) of phonon lines. In the polaron problem, every SCBA diagram corresponds to a unique sequence that records the number of phonons above each propagator. This uniqueness arises from the SCBA rule, which states that phonon lines never cross. The sequence contains exactly the information needed to reconstruct a Dyck path: each up or down step adjusts the path's height, which is all that's needed to fully describe the path. This establishes a clear one-to-one correspondence between Dyck paths and SCBA diagrams, shown together with Dyck paths in Eqs.~(\ref{triv}-\ref{22}).}

\begin{equation}
\begin{aligned}
&a_{0}a_{1}\cdot{}\left(b_{2}a_{1}b_{2}a_{1}+\,b_{2}a_{1}a_{2}b_{3}+\,a_{2}b_{3}b_{2}a_{1}+\,a_{2}b_{3}a_{2}b_{3}+a_{2}a_{3}b_{4}b_{3}\right)\cdot{}b_{2}b_1\\
&\equiv{}
\begin{tikzpicture}[scale=0.7, baseline={([yshift=0.3ex]current bounding box.center)}]
		\fill[red!30] (0,0) -- (1,1) -- (2,2) -- (3,1) -- (4,2) -- (5, 1) -- (6, 2) -- (7, 1) -- (8, 0);
		\draw[ultra thick, black] (0,0) -- (1,1) -- (2,2) -- (3,1) -- (4,2) -- (5, 1) -- (6, 2) -- (7, 1) -- (8, 0);

         \draw[thin, dashed] node at (-0.33,0.1) {0} (0,0)  -- (8,0) ;
		\draw[thin, dashed] node at (-0.33,1) {1}  (0,1) -- (8,1);
		\draw[thin, dashed] node at (-0.33,2) {2}  (0,2) -- (8,2);
        \draw[thin, dashed] node at (-0.33,3) {3}  (0,3) -- (8,3);
        \draw[thin, dashed] node at (-0.33,4) {4}  (0,4) -- (8,4);
\end{tikzpicture}
\quad\left[\;
    \begin{tikzpicture}[scale=0.5, baseline={([yshift=0.3ex]current bounding box.center)}]
  \begin{feynman}
    \vertex (i) at (-4,0);
    \vertex (o) at (-3.2,0);
    \vertex (a) at (-1.6,0);
    \vertex (b) at (-0.8,0);
    \vertex (f) at (0.8,0);
    \vertex (g) at (1.6,0);
    \vertex (h) at (3.2,0);
    \vertex (k) at (4,0);
    \diagram*{
      (i) --[scalar, half left] (k);
      (o) --[scalar, half left] (a);
      (b) --[scalar, half left] (f);
      (g) --[scalar, half left] (h);
      (i) --[fermion2] (o);
      (o) --[fermion2] (a);
      (a) --[fermion2] (b);
      (b) --[fermion2] (f);
      (f) --[fermion2] (g);
      (g) --[fermion2] (h);
      (h) --[fermion2] (k);
    };
  \end{feynman}
\end{tikzpicture}\;\right] \\
&
+
\begin{tikzpicture}[scale=0.7, baseline={([yshift=0.3ex]current bounding box.center)}]
		\fill[red!30] (0,0) -- (1,1) -- (2,2) -- (3,1) -- (4,2) -- (5, 3) -- (6, 2) -- (7, 1) -- (8, 0);
		\draw[ultra thick, black] (0,0) -- (1,1) -- (2,2) -- (3,1) -- (4,2) -- (5, 3) -- (6, 2) -- (7, 1) -- (8, 0);

         \draw[thin, dashed] node at (-0.33,0.1) {0} (0,0)  -- (8,0) ;
		\draw[thin, dashed] node at (-0.33,1) {1}  (0,1) -- (8,1);
		\draw[thin, dashed] node at (-0.33,2) {2}  (0,2) -- (8,2);
        \draw[thin, dashed] node at (-0.33,3) {3}  (0,3) -- (8,3);
        \draw[thin, dashed] node at (-0.33,4) {4}  (0,4) -- (8,4);
\end{tikzpicture}\quad\left[\;
    \begin{tikzpicture}[scale=0.5, baseline={([yshift=0.3ex]current bounding box.center)}]
  \begin{feynman}
    \vertex (i) at (-4,0);
    \vertex (o) at (-3.2,0);
    \vertex (a) at (-1.6,0);
    \vertex (b) at (-0.8,0);
    \vertex (f) at (0.4,0);
    \vertex (g) at (2,0);
    \vertex (h) at (3.2,0);
    \vertex (k) at (4,0);
    \diagram*{
      (i) --[scalar, half left] (k);
      (o) --[scalar, half left] (a);
      (b) --[scalar, half left] (h);
      (f) --[scalar, half left] (g);
      (i) --[fermion2] (o);
      (o) --[fermion2] (a);
      (a) --[fermion2] (b);
      (b) --[fermion2] (f);
      (f) --[fermion2] (g);
      (g) --[fermion2] (h);
      (h) --[fermion2] (k);
    };
  \end{feynman}
\end{tikzpicture}\;\right] \\
&
+
\begin{tikzpicture}[scale=0.7, baseline={([yshift=0.3ex]current bounding box.center)}]
		\fill[red!30] (0,0) -- (1,1) -- (2,2) -- (3,3) -- (4,2) -- (5, 1) -- (6, 2) -- (7, 1) -- (8, 0);
		\draw[ultra thick, black] (0,0) -- (1,1) -- (2,2) -- (3,3) -- (4,2) -- (5, 1) -- (6, 2) -- (7, 1) -- (8, 0);

         \draw[thin, dashed] node at (-0.33,0.1) {0} (0,0)  -- (8,0) ;
		\draw[thin, dashed] node at (-0.33,1) {1}  (0,1) -- (8,1);
		\draw[thin, dashed] node at (-0.33,2) {2}  (0,2) -- (8,2);
        \draw[thin, dashed] node at (-0.33,3) {3}  (0,3) -- (8,3);
        \draw[thin, dashed] node at (-0.33,4) {4}  (0,4) -- (8,4);
\end{tikzpicture}\quad\left[\;
    \begin{tikzpicture}[scale=0.5, baseline={([yshift=0.3ex]current bounding box.center)}]
  \begin{feynman}
    \vertex (i) at (-4,0);
    \vertex (o) at (-3.2,0);
    \vertex (a) at (-2,0);
    \vertex (b) at (-0.4,0);
    \vertex (f) at (0.8,0);
    \vertex (g) at (1.6,0);
    \vertex (h) at (3.2,0);
    \vertex (k) at (4,0);
    \diagram*{
      (i) --[scalar, half left] (k);
      (o) --[scalar, half left] (f);
      (a) --[scalar, half left] (b);
      (g) --[scalar, half left] (h);
      (i) --[fermion2] (o);
      (o) --[fermion2] (a);
      (a) --[fermion2] (b);
      (b) --[fermion2] (f);
      (f) --[fermion2] (g);
      (g) --[fermion2] (h);
      (h) --[fermion2] (k);
    };
  \end{feynman}
\end{tikzpicture}\;\right] \\
&
+
\begin{tikzpicture}[scale=0.7, baseline={([yshift=0.3ex]current bounding box.center)}]
		\fill[red!30](0,0) -- (1,1) -- (2,2) -- (3,3) -- (4,2) -- (5, 3) -- (6, 2) -- (7, 1) -- (8, 0);
		\draw[ultra thick, black] (0,0) -- (1,1) -- (2,2) -- (3,3) -- (4,2) -- (5, 3) -- (6, 2) -- (7, 1) -- (8, 0);

         \draw[thin, dashed] node at (-0.33,0.1) {0} (0,0)  -- (8,0) ;
		\draw[thin, dashed] node at (-0.33,1) {1}  (0,1) -- (8,1);
		\draw[thin, dashed] node at (-0.33,2) {2}  (0,2) -- (8,2);
        \draw[thin, dashed] node at (-0.33,3) {3}  (0,3) -- (8,3);
        \draw[thin, dashed] node at (-0.33,4) {4}  (0,4) -- (8,4);
\end{tikzpicture}\quad\left[\;
    \begin{tikzpicture}[scale=0.5, baseline={([yshift=0.3ex]current bounding box.center)}]
  \begin{feynman}
    \vertex (i) at (-4,0);
    \vertex (o) at (-3.2,0);
    \vertex (a) at (-2.4,0);
    \vertex (b) at (-0.4,0);
    \vertex (f) at (0.4,0);
    \vertex (g) at (2.4,0);
    \vertex (h) at (3.2,0);
    \vertex (k) at (4,0);
    \diagram*{
      (i) --[scalar, half left] (k);
      (o) --[scalar, half left] (h);
      (a) --[scalar, half left] (b);
      (f) --[scalar, half left] (g);
      (i) --[fermion2] (o);
      (o) --[fermion2] (a);
      (a) --[fermion2] (b);
      (b) --[fermion2] (f);
      (f) --[fermion2] (g);
      (g) --[fermion2] (h);
      (h) --[fermion2] (k);
    };
  \end{feynman}
\end{tikzpicture}\;\right] \\
&+
\begin{tikzpicture}[scale=0.7, baseline={([yshift=0.3ex]current bounding box.center)}]
		\fill[red!30] (0,0) -- (1,1) -- (2,2) -- (3,3) -- (4,4) -- (5, 3) -- (6, 2) -- (7, 1) -- (8, 0);
		\draw[ultra thick, black] (0,0) -- (1,1) -- (2,2) -- (3,3) -- (4,4) -- (5, 3) -- (6, 2) -- (7, 1) -- (8, 0);

         \draw[thin, dashed] node at (-0.33,0.1) {0} (0,0)  -- (8,0) ;
		\draw[thin, dashed] node at (-0.33,1) {1}  (0,1) -- (8,1);
		\draw[thin, dashed] node at (-0.33,2) {2}  (0,2) -- (8,2);
        \draw[thin, dashed] node at (-0.33,3) {3}  (0,3) -- (8,3);
        \draw[thin, dashed] node at (-0.33,4) {4}  (0,4) -- (8,4);
\end{tikzpicture}
\quad\left[\;
    \begin{tikzpicture}[scale=0.5, baseline={([yshift=0.3ex]current bounding box.center)}]
  \begin{feynman}
    \vertex (i) at (-4,0);
    \vertex (o) at (-3,0);
    \vertex (a) at (-2,0);
    \vertex (b) at (-1,0);
    \vertex (f) at (1,0);
    \vertex (g) at (2,0);
    \vertex (h) at (3,0);
    \vertex (k) at (4,0);
    \diagram*{
      (i) --[scalar, half left] (k);
      (o) --[scalar, half left] (h);
      (a) --[scalar, half left] (g);
      (b) --[scalar, half left] (f);
      (i) --[fermion2] (o);
      (o) --[fermion2] (a);
      (a) --[fermion2] (b);
      (b) --[fermion2] (f);
      (f) --[fermion2] (g);
      (g) --[fermion2] (h);
      (h) --[fermion2] (k);
    };
  \end{feynman}
\end{tikzpicture}\;\right].
\end{aligned}
\label{26}
\end{equation}

\nver{In particular, to construct an SCBA diagram from a Dyck path, one first draws a fermion line with two external vertices. Internal vertices are then inserted between the external ones, with their number equal to the length of the Dyck path. Proceeding from left to right, a phonon line is created or annihilated at the leftmost available internal vertex for each upward or downward step of the Dyck path, respectively, ensuring that phonon lines do not cross. Finally, the two external vertices are connected by a phonon line.

With this construction, it is straightforward to see that Dyck paths of maximal height, $h=l/2$, correspond to rainbow diagrams. Based on the SCBA diagrams generated from Dyck paths, we can formulate an algorithm that produces the full set of Feynman diagrams, which is presented later in the text.} 

\section{Polynomial expansion of the exact self-energy}
\label{sec4}

\nver{A remarkable property of the self-energy in Eq.~\eqref{SDMFT} is that it incorporates all vertex-correction contributions through simple numerical factors. This becomes apparent by comparing Eq.~\eqref{SDMFT} with the SCBA expression in Eq.~\eqref{SSBOA}. Using Stieltjes-Rogers polynomials, the self-energy in Eq.~\eqref{SDMFT} can be expressed in polynomial form, in direct analogy with the SCBA case. The difference lies in the combinations of numerical factors appearing in the expansion; as we show below, these combinations precisely determine the number of additional diagrams generated when vertex corrections are included.}

Recognizing Eq.~\eqref{SDMFT} as the $\mathcal{S}$-fraction defined by Eq.~\eqref{3}, \nver{with $a_0b_1=G_{0}\left(\omega-\omega_0\right)$}, and the set $\mathfrak{X}$ of coefficients given by

\begin{equation}
a_r = G_{0}\left(\omega-r\cdot{}\omega_{0}\right)\;\;\;,\;\;\;
b_r = r\cdot{}G_{0}\left(\omega-r\cdot{}\omega_{0}\right)\;,
\label{13}
\end{equation}

\noindent Eq.~\eqref{SDMFT} may be written as an infinite series,

\begin{equation}
\bSigma(\omega{}) = g^{2}G\left(\omega-\omega_{0}\right)
\mathcal{S}\left(\mathfrak{X}, g^{2}\right) = g^{2}G(\omega-\omega_{0})\sum_{k = 0}^{+\infty{}}R_{2k}\left(\mathfrak{X}\right)g^{2k}\;,
\label{14}
\end{equation}

\noindent where the Stieltjes-Rogers polynomial of order $2k>0$, $R_{2k}$, is given by Eq.~(\ref{5}). Thus, the self-energy leading-order contribution is given by 

\begin{equation}
\Sigma^{(2)}(\omega)=g^2G(\omega-\omega_{0})\;,
\label{Sigma2}
\end{equation}

\noindent while, for the higher-order $n=2k+2$ contributions, one gets

\begin{eqnarray}
\bSigma^{(n)}(\omega)&=&g^{n}G(\omega-\omega_{0})
\sum_{\substack{h=0 \\ m_0 + \ldots + m_h = n/2-2}}^{n/2-2}\,\Biggl(\,\prod_{j = 0}^{h-1}\binom{m_{j}+m_{j+1}-1}{m_j-1}\,\Biggl)
\nonumber\\&\times&
\Biggl(\,\prod_{r = 0}^{h}(r+2)^{m_{j}}G^{m_j}(\omega-(r+1)\omega_{0})G^{m_j}(\omega-(r+2)\omega_{0})\,\Biggl)\;.
\label{15}
\end{eqnarray}

\noindent In the SCBA case, $b_r$ in Eq.~\eqref{13} \nver{just needs to be replaced by $b_r=G_{0}\left(\omega-r\cdot{}\omega_{0}\right)$.} 

Each contribution $\bSigma^{(n)}(\omega)$ in Eq.~\eqref{15} corresponds to all the self-energy Feynman diagrams for a given order of perturbation theory. The same kind of derivation may be used to express the exact electron propagator $\bG(\omega)$ in terms of Stieltjes-Rogers polynomials as well. The difference is that in the latter case, the diagrammatic expansion involves all the \nver{self-energy contributions to the electron propagator, reducible and irreducible.}

\subsection{Low-order vertex corrections}

\nver{Using Eq.~\eqref{15}, the 2-nd and 4-th order contributions to the exact self-energy in Eq.~\eqref{SDMFT} are obtained in the following form,}

\begin{equation}
\begin{aligned}
\Sigma^{(2)}\left(\omega{},u\in{}\mathcal{P}\cap{}\mathfrak{X}^{*}_{1}\right) &= \delta_{u,\,a_{0}b_{1}}\cdot{}g^{2}\,G_{0}(\omega{}-\omega_{0})
\end{aligned}
\label{20a}
\end{equation}
\begin{equation}
\begin{aligned}
  \Sigma^{(2)}\left(\omega{},a_{0}b_{1}\right)&\equiv{}
\vcenter{\hbox{\begin{tikzpicture}
\end{tikzpicture}}}
\vcenter{\hbox{\begin{tikzpicture}
  \begin{feynman}
    \vertex(i);
    \vertex [right=2.0cm of i] (o);
    \diagram*{
      (i) --[scalar, half left] (o);
      (i) --[fermion2] (o);
    };
  \end{feynman}
\end{tikzpicture}}}\;. 
\end{aligned}
\label{21a}
\end{equation}

\begin{equation}
\begin{aligned}
\Sigma^{(4)}\left(\omega{},u\in{}\mathcal{P}\cap{}\mathfrak{X}^{*}_{2}\right) &= \delta_{u,\,a_{0}a_1b_2b_{1}}\cdot{}2g^{4}\,G_{0}^{2}(\omega{}-\omega_{0})G_{0}(\omega-2\omega_{0})
\end{aligned}
\label{20}
\end{equation}
\begin{equation}
\begin{aligned}
  \Sigma^{(4)}\left(\omega{},a_{0}a_1b_2b_{1}\right)&\equiv{}
\vcenter{\hbox{\begin{tikzpicture}
\end{tikzpicture}}}
\vcenter{\hbox{\begin{tikzpicture}
  \begin{feynman}
    \vertex(i);
    \vertex [right=0.65cm of i] (o);
    \vertex [right=1.3cm of o] (f);
    \vertex [right=0.65cm of f] (g);
    \diagram*{
      (i) --[scalar, half left] (g);
      (o) --[scalar, half left] (f);
      (i) --[fermion2] (o);
      (o) --[fermion2] (f);
      (f) --[fermion2] (g);
    };
  \end{feynman}
\end{tikzpicture}}}~+~
\vcenter{\hbox{\begin{tikzpicture}
  \begin{feynman}
    \vertex (i);
    \vertex [right=0.65cm of i] (o);
    \vertex [right=1.3cm of o] (f);
    \vertex [right=0.65cm of f] (g);
    \diagram*{
      (i) --[scalar, half left] (f);
      (o) --[red, scalar, half left] (g);
      (i) --[fermion2] (o);
      (o) --[red, fermion2] (f);
      (f) --[red, fermion2] (g);
    };
  \end{feynman}
\end{tikzpicture}}}\;. 
\end{aligned}
\label{21}
\end{equation}

\noindent $\mathfrak{X}_{i}^{*}$ denotes all different strings of all lengths that include all $a$ coefficients up to $a_{i-1}$ and all $b$ coefficients up to $b_i$, given by Eq.~\eqref{13}. The phonon and electron propagators highlighted in red correspond to the leading diagrammatic contribution to the exact vertex function in Fig.~\ref{f1}. In the case of the exact local self-energy in Eq.~\eqref{SDMFT}, the two diagrams in Eq.~\eqref{21} contribute identically, resulting in the factor of 2 in Eq.~\eqref{20}. Similar combinatorial factors appear at higher orders, reflecting the number of distinct Feynman diagrams that yield the same contribution.

Following Eq.~\eqref{15}, the diagrammatic contributions corresponding to the sixth order of perturbation theory are given by

\begin{equation}
\begin{aligned}
  \Sigma^{(6)}\left(\omega{},u\in{}\mathcal{P}\cap{}\mathfrak{X}^{*}_{3}\right) &= \delta_{u,\,a_0a_{1}^{2}b_{2}^{2}b_1}\cdot{}4g^{6}\,G_{0}^{3}(\omega{}-\omega_{0})G_{0}^{2}(\omega-2\omega_{0})\\ &+\delta_{u,\,a_0a_{1}a_{2}b_{3}b_{2}b_1}\cdot{}6g^{6}\,G_{0}^{2}(\omega{}-\omega_{0})G_{0}^{2}(\omega-2\omega_{0})G_{0}(\omega-3\omega_{0})
\end{aligned}
\label{23}
\end{equation}
\begin{equation}
\begin{aligned}
  \Sigma^{(6)}\left(\omega{},a_0a_{1}^{2}b_{2}^{2}b_1\right)&\equiv{}
\vcenter{\hbox{\begin{tikzpicture}
\end{tikzpicture}}}
\vcenter{\hbox{\begin{tikzpicture}
  \begin{feynman}
    \vertex (i_1);
    \vertex [right=0.5cm of i_1] (i_2);
    \vertex [right=1cm of i_2] (i_3);
    \vertex [right=0.5cm of i_3] (i_4);
    \vertex [right=1cm of i_4] (i_5);
    \vertex [right=0.5cm of i_5] (i_6);
    \diagram*{
      (i_1) --[scalar, half left] (i_6);
      (i_2) --[scalar, half left] (i_3);
      (i_4) --[scalar, half left] (i_5);
      (i_1) --[fermion2] (i_2);
      (i_2) --[fermion2] (i_3);
      (i_3) --[fermion2] (i_4);
      (i_4) --[fermion2] (i_5);
      (i_5) --[fermion2] (i_6);
    };
  \end{feynman}
\end{tikzpicture}}}~+~
\vcenter{\hbox{\begin{tikzpicture}
  \begin{feynman}
    \vertex (i_1);
    \vertex [right=0.5cm of i_1] (i_2);
    \vertex [right=1cm of i_2] (i_3);
    \vertex [right=0.5cm of i_3] (i_4);
    \vertex [right=1cm of i_4] (i_5);
    \vertex [right=0.5cm of i_5] (i_6);
    \diagram*{
      (i_1) --[scalar, half left] (i_5);
      (i_2) --[scalar, half left] (i_3);
      (i_4) --[red, scalar, half left] (i_6);
      (i_1) --[fermion2] (i_2);
      (i_2) --[fermion2] (i_3);
      (i_3) --[fermion2] (i_4);
      (i_4) --[red, fermion2] (i_5);
      (i_5) --[red, fermion2] (i_6);
    };
  \end{feynman}
\end{tikzpicture}}}~+~\\
&\quad{}
\vcenter{\hbox{\begin{tikzpicture}
\end{tikzpicture}}}
\vcenter{\hbox{\begin{tikzpicture}
  \begin{feynman}
    \vertex (i_1);
    \vertex [right=0.5cm of i_1] (i_2);
    \vertex [right=1cm of i_2] (i_3);
    \vertex [right=0.5cm of i_3] (i_4);
    \vertex [right=1cm of i_4] (i_5);
    \vertex [right=0.5cm of i_5] (i_6);
    \diagram*{
      (i_1) --[scalar, half left] (i_3);
      (i_2) --[red, scalar, half left] (i_6);
      (i_4) --[red, scalar, half left] (i_5);
      (i_1) --[fermion2] (i_2);
      (i_2) --[red, fermion2] (i_3);
      (i_3) --[red, fermion2] (i_4);
      (i_4) --[red, fermion2] (i_5);
      (i_5) --[red, fermion2] (i_6);
    };
  \end{feynman}
\end{tikzpicture}}}~+~
\vcenter{\hbox{\begin{tikzpicture}
  \begin{feynman}
    \vertex (i_1);
    \vertex [right=0.5cm of i_1] (i_2);
    \vertex [right=1cm of i_2] (i_3);
    \vertex [right=0.5cm of i_3] (i_4);
    \vertex [right=1cm of i_4] (i_5);
    \vertex [right=0.5cm of i_5] (i_6);
    \diagram*{
      (i_1) --[scalar, half left] (i_3);
      (i_2) --[red, scalar, half left] (i_5);
      (i_4) --[red, scalar, half left] (i_6);
      (i_1) --[fermion2] (i_2);
      (i_2) --[red, fermion2] (i_3);
      (i_3) --[red, fermion2] (i_4);
      (i_4) --[red, fermion2] (i_5);
      (i_5) --[red, fermion2] (i_6);
    };
  \end{feynman}
\end{tikzpicture}}}\quad{};
\end{aligned}
\label{24}
\end{equation}
\begin{equation}
\begin{aligned}
  \Sigma^{(6)}\left(\omega{},a_0a_{1}a_{2}b_{3}b_{2}b_1\right)&\equiv{}
\vcenter{\hbox{\begin{tikzpicture}
\end{tikzpicture}}}
\vcenter{\hbox{\begin{tikzpicture}
  \begin{feynman}
    \vertex (i_1);
    \vertex [right=0.56cm of i_1] (i_2);
    \vertex [right=0.56cm of i_2] (i_3);
    \vertex [right=0.65cm of i_3] (i_4);
    \vertex [right=0.56cm of i_4] (i_5);
    \vertex [right=0.56cm of i_5] (i_6);
    \diagram*{
      (i_1) --[scalar, half left] (i_6);
      (i_2) --[scalar, half left] (i_5);
      (i_3) --[scalar, half left] (i_4);
      (i_1) --[fermion2] (i_2);
      (i_2) --[fermion2] (i_3);
      (i_3) --[fermion2] (i_4);
      (i_4) --[fermion2] (i_5);
      (i_5) --[fermion2] (i_6);
    };
  \end{feynman}
\end{tikzpicture}}}~+~
\vcenter{\hbox{\begin{tikzpicture}
  \begin{feynman}
    \vertex (i_1);
    \vertex [right=0.5cm of i_1] (i_2);
    \vertex [right=1cm of i_2] (i_3);
    \vertex [right=0.5cm of i_3] (i_4);
    \vertex [right=1cm of i_4] (i_5);
    \vertex [right=0.5cm of i_5] (i_6);
    \diagram*{
      (i_1) --[scalar, half left] (i_6);
      (i_2) --[scalar, half left] (i_4);
      (i_3) --[blue, scalar, half left] (i_5);
      (i_1) --[fermion2] (i_2);
      (i_2) --[fermion2] (i_3);
      (i_3) --[blue, fermion2] (i_4);
      (i_4) --[blue, fermion2] (i_5);
      (i_5) --[fermion2] (i_6);
    };
  \end{feynman}
\end{tikzpicture}}}~+~\\
&\quad{}
\vcenter{\hbox{\begin{tikzpicture}
\end{tikzpicture}}}
\vcenter{\hbox{\begin{tikzpicture}
  \begin{feynman}
    \vertex (i_1);
    \vertex [right=0.56cm of i_1] (i_2);
    \vertex [right=0.56cm of i_2] (i_3);
    \vertex [right=0.7cm of i_3] (i_4);
    \vertex [right=0.56cm of i_4] (i_5);
    \vertex [right=0.56cm of i_5] (i_6);
    \diagram*{
      (i_1) --[scalar, half left] (i_5);
      (i_2) --[red, scalar, half left] (i_6);
      (i_3) --[red, scalar, half left] (i_4);
      (i_1) --[fermion2] (i_2);
      (i_2) --[red, fermion2] (i_3);
      (i_3) --[red, fermion2] (i_4);
      (i_4) --[red, fermion2] (i_5);
      (i_5) --[red, fermion2] (i_6);
    };
  \end{feynman}
\end{tikzpicture}}}~+~
\vcenter{\hbox{\begin{tikzpicture}
  \begin{feynman}
    \vertex (i_1);
    \vertex [right=0.5cm of i_1] (i_2);
    \vertex [right=1cm of i_2] (i_3);
    \vertex [right=0.5cm of i_3] (i_4);
    \vertex [right=1cm of i_4] (i_5);
    \vertex [right=0.5cm of i_5] (i_6);
    \diagram*{
      (i_1) --[scalar, half left] (i_5);
      (i_2) --[red, scalar, half left] (i_4);
      (i_3) --[red, scalar, half left] (i_6);
      (i_1) --[fermion2] (i_2);
      (i_2) --[red, fermion2] (i_3);
      (i_3) --[red, fermion2] (i_4);
      (i_4) --[red, fermion2] (i_5);
      (i_5) --[red, fermion2] (i_6);
    };
  \end{feynman}
\end{tikzpicture}}}
~+~\\
&\quad{}
\vcenter{\hbox{\begin{tikzpicture}
\end{tikzpicture}}}
\vcenter{\hbox{\begin{tikzpicture}
  \begin{feynman}
    \vertex (i_1);
    \vertex [right=0.56cm of i_1] (i_2);
    \vertex [right=0.56cm of i_2] (i_3);
    \vertex [right=0.7cm of i_3] (i_4);
    \vertex [right=0.56cm of i_4] (i_5);
    \vertex [right=0.56cm of i_5] (i_6);
    \diagram*{
      (i_1) --[scalar, half left] (i_4);
      (i_3) --[red, scalar, half left] (i_6);
      (i_2) --[red, scalar, half left] (i_5);
      (i_1) --[fermion2] (i_2);
      (i_2) --[red, fermion2] (i_3);
      (i_3) --[red, fermion2] (i_4);
      (i_4) --[red, fermion2] (i_5);
      (i_5) --[red, fermion2] (i_6);
    };
  \end{feynman}
\end{tikzpicture}}}~+~
\vcenter{\hbox{\begin{tikzpicture}
  \begin{feynman}
    \vertex (i_1);
    \vertex [right=0.5cm of i_1] (i_2);
    \vertex [right=1cm of i_2] (i_3);
    \vertex [right=0.5cm of i_3] (i_4);
    \vertex [right=1cm of i_4] (i_5);
    \vertex [right=0.5cm of i_5] (i_6);
    \diagram*{
      (i_1) --[scalar, half left] (i_4);
      (i_2) --[red, scalar, half left] (i_6);
      (i_3) --[red, scalar, half left] (i_5);
      (i_1) --[fermion2] (i_2);
      (i_2) --[red, fermion2] (i_3);
      (i_3) --[red, fermion2] (i_4);
      (i_4) --[red, fermion2] (i_5);
      (i_5) --[red, fermion2](i_6);
    };
  \end{feynman}
\end{tikzpicture}}}\;.
\end{aligned}
\label{25}
\end{equation}

\noindent Again, the parts of the diagrams corresponding to the exact vertex part of the diagram in Fig.~\ref{f1} are highlighted in red, while the vertex correction that contributes to the exact Green function is in blue. Just as in the fourth order, the sum of the prefactors of different products of free propagators corresponds to the total number of self-energy diagrams.

\par In principle, all other contributions with vertex corrections derived from a given Dyck path can be obtained by specific permutations of the phonon lines in the corresponding SCBA self-energy diagram. In particular, to generate all self-energy contributions for the polaron problem graphically from a given SCBA diagram, one must perform all permutations of temporally retarded phonon-line vertices such that, after permutation, vertices that were temporally retarded remain temporally retarded relative to their temporally advanced counterparts. However, using Eq.~\eqref{cfSigma} and Fig.~\ref{f2}, there exists an alternative and considerably simpler procedure for generating all self-energy diagrams with vertex corrections. As an intermediate step toward deriving this procedure, we first examine the properties of the exact vertex function. 

\section{Closed-form expressions for the exact vertex function}
\label{sec5}

In general, when the self-energy is momentum-dependent, the Ward-Takahashi identity relates the vertex function to the self-energy in the long-wavelength limit only \cite{Mahan,Engelsberg_1963,Pandey_2024}. However, when the self-energy and the bare propagator are local, this identity applies for all momenta equally, which, for the polaron problem, permits a polynomial expression of the exact vertex function. In particular, we discuss next how the diagrammatic contributions to the vertex function appearing in any order of diagrammatic expansion of the electron self-energy may be obtained from the self-energy diagrams obtained in the previous lower order.

\subsection{Ward-Takahashi identity}

For nondispersive phonons, $\omega_q = \omega_0$, the vertex function may be treated as a function of a single variable, $\omega$. Specifically, with respect to frequency dependence, it suffices to know $\Gamma(\omega - \omega_0, \omega)$ to evaluate the continued fraction in Eq.~\eqref{cfSigma}. With all quantities being local, the exact vertex takes on a particularly simple form. By examining the diagram in Fig.~\ref{f1}, one finds that the exact self-energy in Eq.~\eqref{SDMFT} is given by

\begin{equation}
\underline{\Sigma{}}\left(\omega{}\right) = g^{2}\cdot{}\underline{G}\left(\omega-\omega_0\right)\underline{\Gamma}\left(\omega-\omega_0, \omega{}\right)\;,
\label{31}
\end{equation}

\noindent or,

\begin{equation}
\underline{\Gamma{}}\left(\omega-\omega_0, \omega\right) = g^{-2}\;\underline{\Sigma{}}\left(\omega{}\right)\cdot{}\big(G_0^{-1}\left(\omega-\omega_0\right)-\underline{\Sigma{}}\left(\omega-\omega_0\right)\big)\;.
\label{33}
\end{equation}

\noindent Furthermore, this property must hold order by order in the diagrammatic expansion, which, using Eq.~(\ref{14}), gives 

\begin{equation}
\begin{aligned}
\underline{\Gamma{}}\left(\omega-\omega_0, \omega\right) &= 1+\sum_{k = 1}^{+\infty{}}g^{2k}\,\cdot{}\,\bigg\{R_{2k}\left(\omega\right)-G_{0}\left(\omega-\omega_0\right)G_{0}\left(\omega-2\omega_{0}\right)R_{2k-2}\left(\omega-\omega_0\right)\bigg\}
\\&-G_{0}\left(\omega-\omega_0\right)G_{0}\left(\omega-2\omega_{0}\right)\cdot{}\sum_{k,m = 1}^{+\infty{}}g^{2k+2m}\cdot{}R_{2k}\left(\omega\right)R_{2m-2}\left(\omega-\omega_0\right)\;.
\end{aligned}
\label{34}
\end{equation}

\nver{Alternatively, when the exact self-energy is local, the exact vertex may be derived from the Ward-Takahashi identity \cite{Mahan},

\begin{equation}
\underline{\Gamma}\left(\omega-\omega', \omega\right) = 1+\frac{1}{\omega'}\big(\underline{\Sigma}\left(\omega-\omega'\right)-\underline{\Sigma}\left(\omega\right)\big)\;,
\label{36}
\end{equation}

\noindent valid for any $\omega'$. Equation~\eqref{36} can be interpreted as a finite-difference representation of the derivative of the self-energy with respect to an external frequency. In this sense, the vertex function represents the response of the self-energy to an external perturbation. Correspondingly, the diagrammatic construction of the vertex function can be understood as inserting a vertex along electron-propagator lines within the self-energy diagrams.

In particular, by setting $\omega'=\omega_0$, and in addition to Eq.~\eqref{34}, one obtains from Eq.~\eqref{14} an additional expression for the vertex function appearing in Eq.~\eqref{31},

\begin{equation}
\underline{\Gamma}\left(\omega-\omega_0, \omega\right) = 1+\frac{g^{2}}{\omega_0}\sum_{k = 0}^{+\infty{}}g^{2k}\bigg\{G_{0}\left(\omega-2\omega_0\right)R_{2k}\left(\omega-\omega_0\right)-G_{0}\left(\omega-\omega_0\right)R_{2k}\left(\omega\right)\bigg\}\;.
\label{37}
\end{equation}

\noindent The advantage of this latter form is that it admits a direct diagrammatic interpretation. Let us consider the case in which the bare electron propagator is given by a simple pole. Then, the self-energy in Eq.~\eqref{14} involves products of poles. Starting from the mathematical identity proven in Appendix~\ref{App1},

\begin{equation}
\begin{aligned}
    &\,\,\,\,\,\,\,\,\,\,\,\,\,\,\,\prod_{n=1}^{l}\,\frac{1}{(\omega-\omega_{n}-\omega')}\,\,-\,\,\prod_{n=1}^{l}\,\frac{1}{(\omega-\omega_{n})} = \\\sum_{i=1}^{l}\,\,\,&\frac{\omega'}{(\prod_{n<i}(\omega-\omega_n))(\omega-\omega_i)(\omega-\omega_i-\omega')\prod_{n'>i}(\omega-\omega_{n'}-\omega')}\;,
\end{aligned}
\label{WTid}
\end{equation}

\noindent the left-hand side (LHS) of Eq.~\eqref{WTid} reflects the structure of the Ward-Takahashi identity on the right-hand side (RHS) of Eq.~\eqref{36}. In Eq.~\eqref{WTid}, frequencies $\omega_n$ characterize the bare electron propagators shifted by phonon excitation energies, $\omega_n=\varepsilon_k+\sum_q \omega_q$ as they appear in the self-energy diagrams. For the Holstein model $\omega_q=\omega_0$.  On the other hand, the RHS of Eq.~\eqref{WTid} reflects the diagrammatic content of the vertex function. That is, by summing over $i$, one obtains all vertex function contributions corresponding to a given self-energy diagram that is represented by the LHS of Eq.~\eqref{WTid}. All bare electron propagators for $n<i$ remain unchanged, while all those for $n>i$ acquire a shift of $\omega'$. Thus, the self-energy contribution is split into two parts, just as would happen by inserting the bare vertex with one additional phonon excitation propagating to the left/right from this additional vertex. In other words, diagrammatically, any vertex function correction may be generated by inserting the bare vertex along one of the bare electron propagators appearing in the self-energy diagram, drawing the phonon line to the left/right from this additional vertex. This way of obtaining the vertex function diagrammatically has been noted previously \cite{BCS}. Here, this property is explicitly derived for the self-energy given by Eq.~\eqref{14}. The structure of the vertex function is governed by the diagrammatic topology and is therefore general, independent of the specific form of the self-energy.}

As an illustration of how the vertex function corrections may be generated, we start from the leading self-energy diagram. In particular, there is just one vertex function correction that may be obtained by inserting the bare vertex,

\begin{equation}
\begin{aligned}
  \Gamma^{(2)}\left(\omega{}-\omega_0,\omega\right)&\equiv{}\quad{}
\vcenter{\hbox{\begin{tikzpicture}
\end{tikzpicture}}}
\vcenter{\hbox{\begin{tikzpicture}
  \begin{feynman}
    \vertex (i_l) at (-1, 0);
    \vertex (i_r) at (1, 0);
    \vertex (i_mid) at (0, 0);
    \vertex (i_down) at (0, -0.75);
    \diagram*{
      (i_l) --[scalar, half left] (i_r);
      (i_l) --[fermion2] (i_mid);
      (i_mid) --[fermion2] (i_r);
      (i_mid) --[scalar] (i_down);
    };
  \end{feynman}
\end{tikzpicture}}}\quad{}.
\end{aligned}
\label{39}
\end{equation}

\noindent The leading vertex function correction in the local limit is thus given by

\begin{equation}
\Gamma^{(2)}\left(\omega-\omega_0, \omega\right) = g^2G_{0}\left(\omega-\omega_0\right)G_{0}\left(\omega-2\omega_0\right)\;.
\label{40}
\end{equation}

\noindent By closing the external phonon leg onto the incoming fermion leg, and by cutting the remaining external electron leg, one obtains the expression for the second self-energy diagram in Eq.~\eqref{21} with the vertex correction,

\begin{equation}
g^4G_{0}^{2}\left(\omega-\omega_0\right)G_{0}\left(\omega-2\omega_0\right) = g^2G_{0}\left(\omega-\omega_0\right)\cdot{}\Gamma^{(2)}\left(\omega-\omega_0, \omega\right)\;.
\label{41}
\end{equation}

In the next order of the expansion, from the two self-energy diagrams in Eq.~(\ref{21}), one gets all the vertex function corrections in the fourth order, 

\begin{equation}
\begin{aligned}
  \Gamma^{(4)}\left(\omega{}-\omega_0, \omega\right)&\equiv{}\quad{}
\vcenter{\hbox{\begin{tikzpicture}
\end{tikzpicture}}}
\vcenter{\hbox{\begin{tikzpicture}
  \begin{feynman}
    \vertex (i_1) at (-1.7, 0);
    \vertex [right=0.6cm of i_1] (i_2);
    \vertex [right=0.9cm of i_2] (i_3);
    \vertex [right=0.6cm of i_3] (i_4);
    \vertex (i_down) at (0.4, -0.85);
    \vertex [right=0.6cm of i_4] (i_5);
    \diagram*{
      (i_1) --[scalar, half left] (i_5);
      (i_2) --[scalar, half left] (i_3);
      (i_1) --[fermion2] (i_2);
      (i_2) --[fermion2] (i_3);
      (i_3) --[fermion2] (i_4);
      (i_4) --[fermion2] (i_5);
      (i_4) --[scalar] (i_down);
    };
  \end{feynman}
\end{tikzpicture}}}~+~
\vcenter{\hbox{\begin{tikzpicture}
  \begin{feynman}
    \vertex(i) at (-2.7, 0);
    \vertex [right=0.65cm of i] (o);
    \vertex [right=0.7cm of o] (ver);
    \vertex [right=0.7cm of ver] (f);
    \vertex [below = 0.85cm of ver] (ver_dwn);
    \vertex [right=0.65cm of f] (g);
    \diagram*{
      (i) --[scalar, half left] (g);
      (o) --[scalar, half left] (f);
      (i) --[fermion2] (o);
      (o) --[fermion2] (ver);
      (ver) --[fermion2] (f);
      (f) --[fermion2] (g);
      (ver) --[scalar] (ver_dwn);
    };
  \end{feynman}
\end{tikzpicture}}}~+~
\vcenter{\hbox{\begin{tikzpicture}
  \begin{feynman}
    \vertex (i_1) at (-1.7, 0);
    \vertex [right=0.6cm of i_1] (i_4);
    \vertex [right=0.7cm of i_4] (i_2);
    \vertex [right=0.9cm of i_2] (i_3);
    \vertex (i_down) at (-1.1, -0.85);
    \vertex [right=0.6cm of i_3] (i_5);
    \diagram*{
      (i_1) --[scalar, half left] (i_5);
      (i_2) --[scalar, half left] (i_3);
      (i_1) --[fermion2] (i_4);
      (i_4) --[fermion2] (i_2);
      (i_2) --[fermion2] (i_3);
      (i_3) --[fermion2] (i_5);
      (i_4) --[scalar] (i_down);
    };
  \end{feynman}
\end{tikzpicture}}}\\\,\\
&~+~\;
\vcenter{\hbox{\begin{tikzpicture}
  \begin{feynman}
    \vertex (i);
    \vertex [right=0.675cm of i] (o);
    \vertex [right=0.675cm of o] (f);
    \vertex [right=1cm of f] (ver);
    \vertex (ver_dwn) at (0.675, -0.85);
    \vertex [right=0.35cm of ver] (g);
    \diagram*{
      (i) --[scalar, half left] (ver);
      (f) --[scalar, half left] (g);
      (i) --[fermion2] (o);
      (o) --[fermion2] (f);
      (f) --[fermion2] (ver);
      (ver) --[fermion2] (g);
      (o) --[scalar] (ver_dwn);
    };
  \end{feynman}
\end{tikzpicture}}}~+~
\vcenter{\hbox{\begin{tikzpicture}
  \begin{feynman}
    \vertex (i) at (-2.85, 0);
    \vertex [right=0.75cm of i] (o);
    \vertex [right=0.85cm of o] (ver);
    \vertex [below = 0.85cm of ver] (ver_dwn);
    \vertex [right=0.85cm of ver] (f);
    \vertex [right=0.75cm of f] (g);
    \diagram*{
      (i) --[scalar, half left] (f);
      (o) --[scalar, half left] (g);
      (i) --[fermion2] (o);
      (o) --[fermion2] (ver);
      (ver) --[fermion2] (f);
      (f) --[fermion2] (g);
      (ver) --[scalar] (ver_dwn);
    };
  \end{feynman}
\end{tikzpicture}}}~+~
\vcenter{\hbox{\begin{tikzpicture}
  \begin{feynman}
    \vertex (i);
    \vertex [right=0.35cm of i] (o);
    \vertex [right=1cm of o] (f);
    \vertex [right=0.675cm of f] (ver);
    \vertex (ver_dwn) at (2.025, -0.85);
    \vertex [right=0.675cm of ver] (g);
    \diagram*{
      (i) --[scalar, half left] (f);
      (o) --[scalar, half left] (g);
      (i) --[fermion2] (o);
      (o) --[fermion2] (f);
      (f) --[fermion2] (ver);
      (ver) --[fermion2] (g);
      (ver) --[scalar] (ver_dwn);
    };
  \end{feynman}
\end{tikzpicture}}}\;.
\end{aligned}
\label{43}
\end{equation}

\noindent In the local limit, this gives

\begin{equation}
\begin{aligned}
\Gamma^{(4)}\left(\omega-\omega_0, \omega\right)& = 2g^{4}\cdot{}G_{0}^{2}\left(\omega-\omega_0\right)G_{0}^{2}\left(\omega-2\omega_0\right)
\\&+4g^{4}\cdot{}G_{0}\left(\omega-\omega_0\right)G_{0}^{2}\left(\omega-2\omega_0\right)G_{0}\left(\omega-3\omega_0\right)\quad{}.
\end{aligned}
\label{42}
\end{equation}

From Eq.~(\ref{21}), one sees that the self-energy corrections, which were used to construct the vertex function corrections in Eq.~(\ref{43}), may already contain vertex corrections of the lower order. This illustrates, order by order, the diagrammatic content of the Ward-Takahashi identity in Eq.~\eqref{31}. In particular, checking the vertex corrections in red in Eqs.~\eqref{24} and \eqref{25}, one recognizes all the contributions in Eqs.~\eqref{39} and~\eqref{43}. 

\nver{The present discussion shows that all vertex function corrections of a given order $n$ can be obtained by the insertion of the bare vertex along each electron propagator within all self-energy diagrams of the same order. To get all the self-energy diagrams in the next order $n+2$, it is sufficient to consider Fig.~\ref{f2}. That is, one needs to identify all SCBA diagrams up to $n+2$, and insert the vertex function contributions of lower order in such a way that the total order of the resulting self-energy diagrams equals $n+2$.}

\section{Generating Feynman diagrams from Dyck paths\label{SecSS}}

Based on the considerations developed so far, we argue that all the Feynman diagrams appearing in the diagrammatic expansion of the self-energy may be obtained from the Dyck paths \nver{in a very simple way}. The prescribed procedure is a recursive approach where, starting from the leading self-energy diagram, we climb one order at a time, generating all the other diagrammatic contributions to the vertex function and the self-energy in four steps, as schematically depicted in Fig.~\ref{fig:square}. 

\begin{enumerate}[label={}, leftmargin=2\parindent]
        \item \textbf{First step:} For a given order of perturbative expansion $n$, $n\geq2$, derive all the Dyck paths of length $n$. From these paths, draw all the corresponding Feynman SCBA diagrams. 
        
        \item \textbf{Second step:} For all the SCBA diagrams obtained in previous lower orders, replace the bare vertices in the SCBA diagrams with diagrammatic contributions to the vertex, ensuring that the total order of the Feynman self-energy diagrams obtained in this way is $n$. The vertex function corrections have to be inserted according to Fig.~\ref{f2}, replacing the left or right bare vertices in self-energy insertions, but not both of them.
        
        \item \textbf{Third step:} From all the self-energy diagrams obtained in the $n$-th order generate the diagrammatic contributions to the vertex function. All these contributions are obtained by inserting one bare vertex along each electron propagator in each self-energy diagram.

        \item \textbf{Fourth step:} Return to the first step and repeat the procedure for the next order in the perturbative expansion $n+2$.
\end{enumerate}

\begin{figure}[btp]
\centering
\begin{tikzpicture}[
    shorten < =  1mm, shorten > = 1mm,
    node distance = 33mm, on grid, auto,
    every path/.style = {bend left, -Latex}
]
\node (A) {Dyck path(n-2)};
\node (B) [right=of A] {SCBA(n)};
\node (C) [below=of A] {$\Gamma^{(n)}$};
\node (D) [right=of C] {$\Sigma^{(n)}$};

\node (AuxRight) [right= 38mm of D] {SCBA(m < n)};
\node (AuxBelow) [below= 17mm of AuxRight] {$\Gamma^{(l_i\, < \,n)}$};
\node (Intersection) [below left=15mm of AuxRight, draw=none] {};  
\node (Inter) [left=13mm of Intersection] {$m+\sum_i l_i = n$};  


\path[->]   
            (A) edge ["$\textbf{First \,step}$"] (B)
            (B) edge     (D)
            (B) edge [dashed, bend left = 45] (AuxRight)
            (D) edge  ["$\textbf{Third\, step}$"]      (C)
            (C) edge ["$\textbf{Fourth\,step}$"] (A)
            (C) edge [dashed, bend right = 80] (AuxBelow)

            (AuxRight) edge (Inter)
            (AuxBelow) edge[bend left=45] (Inter)  
            (Inter) edge [above right] node {$\textbf{Second\, step}$} (D);
\end{tikzpicture}
    \caption{Scheme of the recursive procedure for the generation of self-energy Feynman diagrams from Dyck paths in four steps. }  
    \label{fig:square}  
\end{figure}

In Table~\ref{tab:diagrams}, we demonstrate how the prescribed procedure generates all Feynman diagrams up to order $n=6$. Starting with $n=2$, in the first step the above procedure gives one Dyck path and the corresponding SCBA self-energy diagram. For $n=2$, the second step does not apply, and we may move to the third step, in which the leading vertex function correction, denoted in red, is obtained. Returning to the first step, for $n=4$ the Dyck path of length four is obtained together with the corresponding SCBA diagram. Following now the prescription for the second step, the leading vertex function correction is inserted into the single $n=2$ SCBA self-energy diagram, yielding the second 4th-order self-energy diagram, with the inserted vertex correction denoted in red in Table~\ref{tab:diagrams}. The latter contributes to the vertex function in Fig.~\ref{f1}. Together with the single $n=4$ SCBA diagram obtained in the first step, this reproduces the two existing 4th-order self-energy diagrams. In the third step, we generate all the 4th-order vertex function corrections from the two 4th-order self-energy diagrams, which are now highlighted in blue. Continuing the procedure further for $n=6$, for example, one may check that these blue vertex function correction diagrams, when inserted into the lowest $n=2$ SCBA self-energy diagram, reproduce the six out of ten 6th-order self-energy diagrams, whose vertex correction parts to the vertex function are similarly denoted in blue in Table~\ref{tab:diagrams}. Two of the ten 6th-order self-energy diagrams are just the SCBA diagrams corresponding to the two Dyck paths of length six. As prescribed in the second step, the remaining two 6th-order self-energy diagrams are obtained by inserting the leading vertex function correction, denoted in red in Table~\ref{tab:diagrams}, into the only existing $n=4$ SCBA diagram. One of these two vertex corrections contributes to the vertex function in Fig.~\ref{f1}, whereas the second contributes to the exact propagator.

The described procedure is intriguing for its combinatorial elegance and for its use of Dyck paths, whose cardinality is drastically smaller than the total number of self-energy diagrams. It provides a straightforward algorithm for generating all self-energy contributions for the polaron problem.

\begin{table}[h]
    \centering
    \renewcommand{\arraystretch}{1.3} 
    \begin{tabular}{c||cc||c||c}
        & $\mathcal{D}^{(n-2)}$ & $\Sigma^{(n)}_{SCBA}$  & $\Sigma^{(n)}$ & $\Gamma^{(n)}$  \\
        \hline
        \hline
        $n = 2$    & 	\begin{tikzpicture}[scale=0.15]
		\fill[red!30] (0,0) -- (1,1) -- (2,0);
        \draw[ultra thick, black] (0,0) -- (1,1) -- (2,0);
	\end{tikzpicture}  & \begin{tikzpicture}[scale=0.6]
  \begin{feynman}
    \vertex (i_l) at (-1, 0);
    \vertex (i_r) at (1, 0);
    \diagram*{
      (i_l) --[scalar, half left] (i_r);
      (i_l) --[plain] (i_r);
    };
  \end{feynman}
\end{tikzpicture}   & \begin{tikzpicture}[scale=0.6]
  \begin{feynman}
    \vertex (i_l) at (-1, 0);
    \vertex (i_r) at (1, 0);
    \diagram*{
      (i_l) --[scalar, half left] (i_r);
      (i_l) --[plain] (i_r);
    };
  \end{feynman}
\end{tikzpicture}  & \begin{tikzpicture}[scale=0.6]
  \begin{feynman}
    \vertex (i_l) at (-1, 0);
    \vertex (i_r) at (1, 0);
    \vertex (i_mid) at (0, 0);
    \vertex (i_down) at (0, -0.75);
    \diagram*{
      (i_l) --[red, scalar, half left] (i_r);
      (i_l) --[red, plain] (i_mid);
      (i_mid) --[red, plain] (i_r);
      (i_mid) --[red, scalar] (i_down);
    };
  \end{feynman}
\end{tikzpicture}  \\
        \hline\hline
        $n = 4$    & \begin{tikzpicture}[scale=0.15]

		
		\fill[red!30] (0,0) -- (1,1) -- (2,2) -- (3, 1) -- (4, 0);
		\draw[ultra thick, black] (0,0) -- (1,1) -- (2,2) -- (3, 1) -- (4, 0);
\end{tikzpicture}  & \begin{tikzpicture}[scale=0.6]
  \begin{feynman}
    \vertex (i) at (-1,0);
    \vertex (o) at (-0.5,0);
    \vertex (f) at (0.5,0);
    \vertex (g) at (1,0);
    \diagram*{
      (i) --[scalar, half left] (g);
      (o) --[scalar, half left] (f);
      (i) --[plain] (o);
      (o) --[plain] (f);
      (f) --[plain] (g);
    };
  \end{feynman}
\end{tikzpicture}  &  \makecell{\begin{tikzpicture}[scale=0.6]
  \begin{feynman}
    \vertex (i) at (-1,0);
    \vertex (o) at (-0.5,0);
    \vertex (f) at (0.5,0);
    \vertex (g) at (1,0);
    \diagram*{
      (i) --[scalar, half left] (g);
      (o) --[scalar, half left] (f);
      (i) --[plain] (o);
      (o) --[plain] (f);
      (f) --[plain] (g);
    };
  \end{feynman}
\end{tikzpicture} \\ \begin{tikzpicture}[scale=0.6]
  \begin{feynman}
    \vertex (i) at (-1,0);
    \vertex (o) at (-0.5,0);
    \vertex (f) at (0.5,0);
    \vertex (g) at (1,0);
    \diagram*{
      (i) --[scalar, half left] (f);
      (o) --[red, scalar, half left] (g);
      (i) --[plain] (o);
      (o) --[red, plain] (f);
      (f) --[red, plain] (g);
    };
  \end{feynman}
\end{tikzpicture} }  & \makecell{
\begin{tikzpicture}[scale=0.6]
  \begin{feynman}
    \vertex (i_1) at (-1,0);
    \vertex (i_2) at (-0.75,0);
    \vertex (i_3) at (-0.5,0);
    \vertex (i_4) at (0.5,0);
    \vertex (i_5) at (1,0);
    \vertex (i_down) at (-0.75,-0.5);
    \diagram*{
      (i_1) --[blue, scalar, half left] (i_5);
      (i_3) --[blue, scalar, half left] (i_4);
      (i_1) --[blue, plain] (i_2);
      (i_2) --[blue, plain] (i_3);
      (i_3) --[blue, plain] (i_4);
      (i_4) --[blue, plain] (i_5);
      (i_2) --[blue, scalar] (i_down);
    };
  \end{feynman}
\end{tikzpicture}  \begin{tikzpicture}[scale=0.6]
  \begin{feynman}
    \vertex (i_1) at (-1,0);
    \vertex (i_2) at (-0.5,0);
    \vertex (i_3) at (0.0,0);
    \vertex (i_4) at (0.5,0);
    \vertex (i_5) at (1,0);
    \vertex (i_down) at (0.0,-0.5);
    \diagram*{
      (i_1) --[blue, scalar, half left] (i_5);
      (i_2) --[blue, scalar, half left] (i_4);
      (i_1) --[blue, plain] (i_2);
      (i_2) --[blue, plain] (i_3);
      (i_3) --[blue, plain] (i_4);
      (i_4) --[blue, plain] (i_5);
      (i_3) --[blue, scalar] (i_down);
    };
  \end{feynman}
\end{tikzpicture}
\begin{tikzpicture}[scale=0.6]
  \begin{feynman}
    \vertex (i_1) at (-1,0);
    \vertex (i_2) at (-0.5,0);
    \vertex (i_3) at (0.5,0);
    \vertex (i_4) at (0.75,0);
    \vertex (i_5) at (1,0);
    \vertex (i_down) at (0.75,-0.5);
    \diagram*{
      (i_1) --[blue, scalar, half left] (i_5);
      (i_2) --[blue, scalar, half left] (i_3);
      (i_1) --[blue, plain] (i_2);
      (i_2) --[blue, plain] (i_3);
      (i_3) --[blue, plain] (i_4);
      (i_4) --[blue, plain] (i_5);
      (i_4) --[blue, scalar] (i_down);
    };
  \end{feynman}
\end{tikzpicture} \\ 
\begin{tikzpicture}[scale=0.6]
  \begin{feynman}
    \vertex (i_1) at (-1,0);
    \vertex (i_2) at (-0.75,0);
    \vertex (i_3) at (-0.5,0);
    \vertex (i_4) at (0.5,0);
    \vertex (i_5) at (1,0);
    \vertex (i_down) at (-0.75,-0.5);
    \diagram*{
      (i_1) --[blue, scalar, half left] (i_4);
      (i_3) --[blue, scalar, half left] (i_5);
      (i_1) --[blue, plain] (i_2);
      (i_2) --[blue, plain] (i_3);
      (i_3) --[blue, plain] (i_4);
      (i_4) --[blue, plain] (i_5);
      (i_2) --[blue, scalar] (i_down);
    };
  \end{feynman}
\end{tikzpicture}
\begin{tikzpicture}[scale=0.6]
  \begin{feynman}
    \vertex (i_1) at (-1,0);
    \vertex (i_2) at (-0.5,0);
    \vertex (i_3) at (0.0,0);
    \vertex (i_4) at (0.5,0);
    \vertex (i_5) at (1,0);
    \vertex (i_down) at (0.0,-0.5);
    \diagram*{
      (i_1) --[blue, scalar, half left] (i_4);
      (i_2) --[blue, scalar, half left] (i_5);
      (i_1) --[blue, plain] (i_2);
      (i_2) --[blue, plain] (i_3);
      (i_3) --[blue, plain] (i_4);
      (i_4) --[blue, plain] (i_5);
      (i_3) --[blue, scalar] (i_down);
    };
  \end{feynman}
\end{tikzpicture}
\begin{tikzpicture}[scale=0.6]
  \begin{feynman}
    \vertex (i_1) at (-1,0);
    \vertex (i_2) at (-0.5,0);
    \vertex (i_3) at (0.5,0);
    \vertex (i_4) at (0.75,0);
    \vertex (i_5) at (1,0);
    \vertex (i_down) at (0.75,-0.5);
    \diagram*{
      (i_1) --[blue, scalar, half left] (i_3);
      (i_2) --[blue, scalar, half left] (i_5);
      (i_1) --[blue, plain] (i_2);
      (i_2) --[blue, plain] (i_3);
      (i_3) --[blue, plain] (i_4);
      (i_4) --[blue, plain] (i_5);
      (i_4) --[blue, scalar] (i_down);
    };
  \end{feynman}
\end{tikzpicture}}
\\
        \hline\hline
        $n = 6$    & \begin{tikzpicture}[scale=0.15]
		
		\fill[red!30] (0,0) -- (1,1) -- (2,2) -- (3,1) -- (4,2) -- (5, 1) -- (6, 0);
		\draw[ultra thick, black] (0,0) -- (1,1) -- (2,2) -- (3,1) -- (4,2) -- (5, 1) -- (6, 0);

\end{tikzpicture}

\begin{tikzpicture}[scale = 0.15]
		
		\fill[red!30] (0,0) -- (1,1) -- (2,2) -- (3,3) -- (4,2) -- (5, 1) -- (6, 0);
		\draw[ultra thick, black] (0,0) -- (1,1) -- (2,2) -- (3,3) -- (4,2) -- (5, 1) -- (6, 0);

\end{tikzpicture}  & 
\begin{tikzpicture}[scale=0.6]
  \begin{feynman}
    \vertex (i) at (-1,0);
    \vertex (o) at (-0.75,0);
    \vertex (a) at (-0.075,0);
    \vertex (b) at (0.075,0);
    \vertex (f) at (0.75,0);
    \vertex (g) at (1,0);
    \diagram*{
      (i) --[scalar, half left] (g);
      (o) --[scalar, half left] (a);
      (b) --[scalar, half left] (f);
      (i) --[plain] (o);
      (o) --[plain] (a);
      (a) --[plain] (b);
      (b) --[plain] (f);
      (f) --[plain] (g);
    };
  \end{feynman}
\end{tikzpicture} 
\begin{tikzpicture}[scale=0.6]
  \begin{feynman}
    \vertex (i) at (-1,0);
    \vertex (o) at (-2/3,0);
    \vertex (a) at (-1/3,0);
    \vertex (b) at (1/3,0);
    \vertex (f) at (2/3,0);
    \vertex (g) at (1,0);
    \diagram*{
      (i) --[scalar, half left] (g);
      (o) --[scalar, half left] (f);
      (a) --[scalar, half left] (b);
      (i) --[plain] (o);
      (o) --[plain] (a);
      (a) --[plain] (b);
      (b) --[plain] (f);
      (f) --[plain] (g);
    };
  \end{feynman}
\end{tikzpicture}  & \makecell{\begin{tikzpicture}[scale=0.6]
  \begin{feynman}
    \vertex (i) at (-1,0);
    \vertex (o) at (-0.75,0);
    \vertex (a) at (-0.075,0);
    \vertex (b) at (0.075,0);
    \vertex (f) at (0.75,0);
    \vertex (g) at (1,0);
    \diagram*{
      (i) --[scalar, half left] (g);
      (o) --[scalar, half left] (a);
      (b) --[scalar, half left] (f);
      (i) --[plain] (o);
      (o) --[plain] (a);
      (a) --[plain] (b);
      (b) --[plain] (f);
      (f) --[plain] (g);
    };
  \end{feynman}
\end{tikzpicture} 
\begin{tikzpicture}[scale=0.6]
  \begin{feynman}
    \vertex (i) at (-1,0);
    \vertex (o) at (-2/3,0);
    \vertex (a) at (-1/3,0);
    \vertex (b) at (1/3,0);
    \vertex (f) at (2/3,0);
    \vertex (g) at (1,0);
    \diagram*{
      (i) --[scalar, half left] (g);
      (o) --[scalar, half left] (f);
      (a) --[scalar, half left] (b);
      (i) --[plain] (o);
      (o) --[plain] (a);
      (a) --[plain] (b);
      (b) --[plain] (f);
      (f) --[plain] (g);
    };
  \end{feynman}
\end{tikzpicture} \\ \begin{tikzpicture}[scale=0.6]
  \begin{feynman}
    \vertex (i) at (-1,0);
    \vertex (o) at (-0.75,0);
    \vertex (a) at (-0.2,0);
    \vertex (b) at (0.2,0);
    \vertex (f) at (0.75,0);
    \vertex (g) at (1,0);
    \diagram*{
      (i) --[scalar, half left] (g);
      (o) --[scalar, half left] (b);
      (a) --[red, scalar, half left] (f);
      (i) --[plain] (o);
      (o) --[plain] (a);
      (a) --[red, plain] (b);
      (b) --[red, plain] (f);
      (f) --[plain] (g);
    };
  \end{feynman}
\end{tikzpicture} 
\begin{tikzpicture}[scale=0.6]
  \begin{feynman}
    \vertex (i) at (-1,0);
    \vertex (o) at (-0.8,0);
    \vertex (a) at (-0.1,0);
    \vertex (b) at (0.3,0);
    \vertex (f) at (0.6,0);
    \vertex (g) at (1,0);
    \diagram*{
      (i) --[scalar, half left] (f);
      (o) --[scalar, half left] (a);
      (b) --[red, scalar, half left] (g);
      (i) --[plain] (o);
      (o) --[plain] (a);
      (a) --[plain] (b);
      (b) --[red, plain] (f);
      (f) --[red, plain] (g);
    };
  \end{feynman}
\end{tikzpicture} \\ \begin{tikzpicture}[scale=0.6]
  \begin{feynman}
    \vertex (i) at (-1,0);
    \vertex (o) at (-0.6,0);
    \vertex (a) at (-0.3,0);
    \vertex (b) at (0.1,0);
    \vertex (f) at (0.8,0);
    \vertex (g) at (1,0);
    \diagram*{
      (o) --[blue, scalar, half left] (g);
      (i) --[scalar, half left] (a);
      (b) --[blue, scalar, half left] (f);
      (i) --[plain] (o);
      (o) --[blue, plain] (a);
      (a) --[blue, plain] (b);
      (b) --[blue, plain] (f);
      (f) --[blue, plain] (g);
    };
  \end{feynman}
\end{tikzpicture} 
\begin{tikzpicture}[scale=0.6]
  \begin{feynman}
    \vertex (i) at (-1,0);
    \vertex (o) at (-2/3,0);
    \vertex (a) at (-1/3,0);
    \vertex (b) at (1/3,0);
    \vertex (f) at (2/3,0);
    \vertex (g) at (1,0);
    \diagram*{
      (i) --[scalar, half left] (b);
      (o) --[blue, scalar, half left] (g);
      (a) --[blue, scalar, half left] (f);
      (i) --[plain] (o);
      (o) --[blue, plain] (a);
      (a) --[blue, plain] (b);
      (b) --[blue, plain] (f);
      (f) --[blue, plain] (g);
    };
  \end{feynman}
\end{tikzpicture} \begin{tikzpicture}[scale=0.6]
  \begin{feynman}
    \vertex (i) at (-1,0);
    \vertex (o) at (-2/3,0);
    \vertex (a) at (-1/3,0);
    \vertex (b) at (1/3,0);
    \vertex (f) at (2/3,0);
    \vertex (g) at (1,0);
    \diagram*{
      (i) --[scalar, half left] (f);
      (o) --[blue, scalar, half left] (g);
      (a) --[blue, scalar, half left] (b);
      (i) --[plain] (o);
      (o) --[blue, plain] (a);
      (a) --[blue, plain] (b);
      (b) --[blue, plain] (f);
      (f) --[blue, plain] (g);
    };
  \end{feynman}
\end{tikzpicture} \\ \begin{tikzpicture}[scale=0.6]
  \begin{feynman}
    \vertex (i) at (-1,0);
    \vertex (o) at (-2/3,0);
    \vertex (a) at (-1/3,0);
    \vertex (b) at (1/3,0);
    \vertex (f) at (2/3,0);
    \vertex (g) at (1,0);
    \diagram*{
      (i) --[scalar, half left] (a);
      (o) --[blue, scalar, half left] (f);
      (b) --[blue, scalar, half left] (g);
      (i) --[plain] (o);
      (o) --[blue, plain] (a);
      (a) --[blue, plain] (b);
      (b) --[blue, plain] (f);
      (f) --[blue, plain] (g);
    };
  \end{feynman}
\end{tikzpicture} 
\begin{tikzpicture}[scale=0.6]
  \begin{feynman}
    \vertex (i) at (-1,0);
    \vertex (o) at (-2/3,0);
    \vertex (a) at (-1/3,0);
    \vertex (b) at (1/3,0);
    \vertex (f) at (2/3,0);
    \vertex (g) at (1,0);
    \diagram*{
      (i) --[scalar, half left] (b);
      (o) --[blue, scalar, half left] (f);
      (a) --[blue, scalar, half left] (g);
      (i) --[plain] (o);
      (o) --[blue, plain] (a);
      (a) --[blue, plain] (b);
      (b) --[blue, plain] (f);
      (f) --[blue, plain] (g);
    };
  \end{feynman}
\end{tikzpicture} \begin{tikzpicture}[scale=0.6]
  \begin{feynman}
    \vertex (i) at (-1,0);
    \vertex (o) at (-2/3,0);
    \vertex (a) at (-1/3,0);
    \vertex (b) at (1/3,0);
    \vertex (f) at (2/3,0);
    \vertex (g) at (1,0);
    \diagram*{
      (i) --[scalar, half left] (f);
      (o) --[blue, scalar, half left] (b);
      (a) --[blue, scalar, half left] (g);
      (i) --[plain] (o);
      (o) --[blue, plain] (a);
      (a) --[blue, plain] (b);
      (b) --[blue, plain] (f);
      (f) --[blue, plain] (g);
    };
  \end{feynman}
\end{tikzpicture}} & \makecell{\begin{tikzpicture}[scale=0.6]
  \begin{feynman}
    \vertex (i) at (-1.,0);
    \vertex (c) at (-0.875, 0);
    \vertex (down) at (-0.875, -0.5);
    \vertex (o) at (-0.75,0);
    \vertex (a) at (-0.075,0);
    \vertex (b) at (0.075,0);
    \vertex (f) at (0.75,0);
    \vertex (g) at (1,0);
    \diagram*{
      (i) --[MatteGold, scalar, half left] (g);
      (o) --[MatteGold, scalar, half left] (a);
      (c) --[MatteGold, scalar] (down);
      (b) --[MatteGold, scalar, half left] (f);
      (i) --[MatteGold, plain] (c);
      (c) --[MatteGold, plain] (o);
      (o) --[MatteGold, plain] (a);
      (a) --[MatteGold, plain] (b);
      (b) --[MatteGold, plain] (f);
      (f) --[MatteGold, plain] (g);
    };
  \end{feynman}
\end{tikzpicture}... \begin{tikzpicture}[scale=0.6]
  \begin{feynman}
    \vertex (i) at (-1.,0);
    \vertex (c) at (-0.80, 0);
    \vertex (down) at (-0.80, -0.5);
    \vertex (o) at (-0.65,0);
    \vertex (a) at (-0.275,0);
    \vertex (b) at (0.275,0);
    \vertex (f) at (0.65,0);
    \vertex (g) at (1,0);
    \diagram*{
      (i) --[MatteGold, scalar, half left] (g);
      (o) --[MatteGold, scalar, half left] (f);
      (c) --[MatteGold, scalar] (down);
      (a) --[MatteGold, scalar, half left] (b);
      (i) --[MatteGold, plain] (c);
      (c) --[MatteGold, plain] (o);
      (o) --[MatteGold, plain] (a);
      (a) --[MatteGold, plain] (b);
      (b) --[MatteGold, plain] (f);
      (f) --[MatteGold, plain] (g);
    };
  \end{feynman}
\end{tikzpicture}...\\
\begin{tikzpicture}[scale=0.6]
  \begin{feynman}
    \vertex (i) at (-1.,0);
    \vertex (c) at (-0.875, 0);
    \vertex (down) at (-0.875, -0.5);
    \vertex (o) at (-0.75,0);
    \vertex (a) at (-0.075,0);
    \vertex (b) at (0.075,0);
    \vertex (f) at (0.75,0);
    \vertex (g) at (1,0);
    \diagram*{
      (i) --[MatteGold, scalar, half left] (g);
      (o) --[MatteGold, scalar, half left] (b);
      (c) --[MatteGold, scalar] (down);
      (a) --[MatteGold, scalar, half left] (f);
      (i) --[MatteGold, plain] (c);
      (c) --[MatteGold, plain] (o);
      (o) --[MatteGold, plain] (a);
      (a) --[MatteGold, plain] (b);
      (b) --[MatteGold, plain] (f);
      (f) --[MatteGold, plain] (g);
    };
  \end{feynman}
\end{tikzpicture}...
\begin{tikzpicture}[scale=0.6]
  \begin{feynman}
    \vertex (i) at (-1.,0);
    \vertex (c) at (-0.875, 0);
    \vertex (down) at (-0.875, -0.5);
    \vertex (o) at (-0.75,0);
    \vertex (a) at (-0.075,0);
    \vertex (b) at (0.075,0);
    \vertex (f) at (0.75,0);
    \vertex (g) at (1,0);
    \diagram*{
      (i) --[MatteGold, scalar, half left] (f);
      (o) --[MatteGold, scalar, half left] (a);
      (c) --[MatteGold, scalar] (down);
      (b) --[MatteGold, scalar, half left] (g);
      (i) --[MatteGold, plain] (c);
      (c) --[MatteGold, plain] (o);
      (o) --[MatteGold, plain] (a);
      (a) --[MatteGold, plain] (b);
      (b) --[MatteGold, plain] (f);
      (f) --[MatteGold, plain] (g);
    };
  \end{feynman}
\end{tikzpicture}...\\\vdots{}\\
}
    \end{tabular}
    \caption{Order-by-order generation of the electron self-energy and vertex function corrections (highlighted in color) from Dyck paths. The vertex corrections contribute to both the exact electron propagator and the vertex function shown in Fig.~\ref{f1}.}
    \label{tab:diagrams}
\end{table}

\nver{\subsection{Finite electron densities}

Although our algorithm is developed here for the polaron problem, it can be readily extended to systems with finite electron densities. In such cases, one must compute not only the electron self-energy and vertex-function contributions, but also include phonon self-energy contributions. For this extension, two additional aspects of diagram construction should be considered.

(i) Regarding phonon self-energy, the leading, second-order phonon self-energy diagram, denoted as $\Pi^{(2)} \equiv \Pi_0$, is a bubble diagram. It involves two bare fermion Green's functions and two bare electron-phonon vertices. All higher-order ($n > 4$), phonon self-energy diagrams may be obtained by dressing both fermion lines and one electron-phonon vertex. Thus, they are constructed from $\Pi_0$, fermion self-energies $\Sigma^{(m_i)}$, and vertex functions $\Gamma^{(l_i)}$ in lower orders, such that the total diagrammatic order is given by  $n=2+\sum_{i}^{}m_i + l_{j}$. Here, the vertex functions $\Gamma^{(l_i)}$ must be inserted at either the rightmost or leftmost bare electron-phonon vertex of the bubble. 

(ii) All fermion self-energy diagrams at a given order $n$ should be generated by combining SCBA diagrams $\Sigma_{\text{SCBA}}^{(m)}$ with phonon self-energies $\Pi^{(k_i)}$ and vertex corrections $\Gamma^{(l_i)}$, such that $n = m + \sum_{i}^{}k_i + \sum_{j}^{}l_j $. In these diagrams, vertex function corrections replace either the left or right bare vertex, but only within self-energy insertions. 

According to the Ward identity, the topology of all vertex corrections $\Gamma$ can again be derived from the fermion self-energy diagrams at the corresponding order. Compared to the polaron case, the main complication in the finite-density extension lies in the combinatorially large number of ways to construct higher-order diagrams from lower-order $\Sigma$, $\Pi$, and $\Gamma$ components. While handling these constructions is beyond the scope of the present framework, it is intended to be addressed in greater detail in future work. 

\color{black}{Finally, we note that, at finite densities, fermionic Green's functions include both forward and backward propagation in time \cite{Marsiglio1992,Dolgov2008,Tupitsyn2016,Krsnik2022,Erhardt2025}. This property is already present at the level of the bare Green's function and is reflected, for example, in the structure of the leading phonon self-energy diagram, representing electron-hole pair excitations.}}

\subsection{Cardinality of diagrams}

\nver{We now return to the polaron problem to count the distinct diagrammatic contributions. Specifically, we derive analytical expressions for the total number of self-energy diagrams, the contributions arising from vertex corrections, and the exact electron propagator.}

\nver{We consider the problem of counting all diagrammatic contributions to the exact electron propagator at the $n$-th order of perturbation theory. This is achieved by reducing the problem step by step to simpler combinatorial ones. The contributions include both reducible and irreducible self-energy diagrams. We consider a set of $n$ vertices and count the number of distinct pairings of two points.} The number of unordered samples without replacement of cardinality two of the set of $n$ vertices is given by a $2$-combination of the set of vertices. With every pair chosen, the cardinality of the set of vertices is subtracted by two until we get down to only two vertices. Every step brings about the $2$-combination factor equal to $\binom{2j-2i}{2}$, where $i$ is the number of steps taken, so the total number of ways would be equal to the multiplication of said binomials,

\begin{equation}
\mathcal{I}_{tot}^{(n)} = \prod_{j = 1}^{n/2}\,\frac{1}{j}\,\,\binom{2j}{2}\quad{}.
\label{44p}
\end{equation}

\noindent The above expression contains another factor equal to the inverse of the number of steps taken $\frac{1}{j}$. Using the associativity of multiplication on the real domain, the factor $\frac{1}{j}$ becomes $\frac{1}{(n/2)!}$. The factor $\frac{1}{(n/2)!}$ comes from the indistinguishability of phonon lines. Thus, for the total number of \nver{diagrammatic contributions to the exact electron propagator,} one gets

\begin{equation}
\mathcal{I}_{tot}^{(n)}=(n-1)!!\;.
\label{45}
\end{equation}

To address the problem of counting exact self-energy diagrams in the $n$-th order, we begin by observing that the prefactor in front of the propagator product specifies the number of diagrams generated by the algebraic expression. Therefore, to determine the total number of self-energy diagrams, it suffices to sum the prefactors in Eq.~(\ref{15}). Specifically, setting $a_i = 1$, and $b_i = i$, for each $i$, the expression for the number of irreducible diagrams in the $(n+2)$-th order is obtained,

\begin{equation}
\mathcal{I}^{(n+2)}_{irr}\left.\right|_{n\geq{}2} = \sum_{\substack{h=0 \\ m_0 + \ldots + m_h = n/2-1}}^{n/2-1}\,\,\Biggl(\,\prod_{j = 0}^{h-1}\binom{i_{j}+i_{j+1}-1}{i_j-1}\,\Biggl)\cdot{}\Biggl(\,\prod_{l = 0}^{h}(l+2)^{i_{l}}\,\Biggl)\;.
\label{46p}
\end{equation}

\noindent The indexing $n+2$ arises from the additional $g^{2}$ in Eqs.~(\ref{14}) and (\ref{15}), so the $n/2$-th term in the summation in Eq.~(\ref{14}) counts the $g^{n+2}$ contribution to the self-energy. Therefore, Eq.~(\ref{46p}) gives an analytical expression for the number of self-energy corrections order by order of perturbation theory. That is, using Eq.~(\ref{46p}), we may reproduce the known series $1,\,2,\,10,\,74,\,706,\,8162\ldots{}$\cite{Goodvin}, which describes the number of irreducible corrections in each order of perturbation starting from $n = 2$, i.e., $g^{2}$ contribution. Now, it is easy to obtain the number of reducible diagrams by subtracting the \nver{cardinality of all diagrammatic contributions to the exact electronic propagator and the cardinality of self-energy diagrams}, $\mathcal{I}_{red}^{(n)}=\mathcal{I}_{tot}^{(n)} - \mathcal{I}^{(n)}_{irr}$.

The number of SCBA diagrams follows from the bijective mapping from the set of labeled Dyck paths to SCBA-approximated irreducible Feynman diagrams, established in Sec.~\ref{SecSS}. Therefore, by counting Dyck paths of length $n$, one obtains the SCBA contributions to the self-energy in the $n$th order as well. The number of Dyck paths is given by the famous Catalan sequence \cite{Stanley}, so given the equivalence we write down the number of SCBA-approximated diagrams as,

\begin{equation}
\mathcal{I}^{(n)}_{\text{SCBA}} = C_{n/2} = \frac{1}{n/2+1}\binom{n}{n/2}\;.
\label{48p}
\end{equation}

As already shown, all the diagrammatic contributions to the vertex function may be generated by inspecting the self-energy diagrams of the same order and by inserting an additional phonon leg. The phonon leg has to be inserted into every free fermionic propagator. The self-energy contribution of the $n$-th order has $n-1$ free fermionic propagators. Thus, one just needs to multiply that by the number of irreducible diagrams to which we can insert the phonon leg.

\begin{equation}
\mathcal{I}^{(n)}_{\textit{ver. corr.}} = (n-1)\cdot{}\mathcal{I}^{(n)}_{\textit{irr}}\;.
\label{n_ver_corr_p}
\end{equation}

\begin{table}[tb!]
\centering
\begin{tabular}{r r r r r r r} 
{\textbf{Order in coupling constant}}	& $g^2$	& $g^4$ & $g^6$ & $g^8$ & $g^{10}$ & $g^{12}$ \\
&&&&&&\\
Cardinality of SCBA	diagrams	& 1 & 1 & 2 & 5 & 14 & 42 \\
$(n/2)!$		& 1 & 2 & 6 & 24 & 120 & 720 \\
Cardinality of $\bSigma$ diagrams		& 1 & 2 & 10 & 74 & 706 & 8162 \\
Cardinality of diagrammatic contributions to $\bGamma$		& 1 & 6 & 50 & 518 & 6354 & 89782 \\
Cardinality of diagrammatic contributions to $\bG$		& 1 & 3 & 15 & 105 & 945 & 10395 

\end{tabular}
\caption{The comparison of the cardinality of Feynman diagrams in each order of perturbative expansion: number of SCBA approximated diagrammatic contributions, $n/2!$ growth, number of $\bSigma$ diagrams, number of diagrammatic contributions to the vertex function, number of diagrammatic contributions to the exact Green function~$\bG$.\label{tab1}}
\label{t1}
\end{table}

Table~\ref{t1} shows the cardinality of the SCBA diagrams, the cardinality of exact self-energy diagrams, and the cardinality of diagrammatic contributions to the vertex function. For comparison of growth rates, a factorial dependence is included as well. From Table~\ref{t1} one sees that the number of SCBA-approximated diagrams, or the number of Dyck paths, grows very slowly compared to all the other quantities. Furthermore, we observe that the cardinality of diagrammatic contributions to the vertex function, which are the primary contributors to the complexity of the polaron problem, grows faster than $(n/2)!$. This highlights the critical role of vertex corrections in determining the computational difficulty of electron-phonon problems.

\section{Stochastic evaluations of diagrams\label{SecBLF}}
\nver{
Stochastic evaluations of diagrammatic expansions are frequently plagued by the sign problem, due to the alternating signs of Feynman diagrams. In such cases, an approach that evaluates all diagrams of a given order at once for a given set of internal variables has a clear advantage over standard schemes, in which each diagram is evaluated separately and the algorithm stochastically jumps from one diagram topology to another. When all diagrams are evaluated for the same internal configuration, their contributions are correlated, so that cancellations between diagrams occur already at the level of individual Monte Carlo samples rather than only after averaging over many updates.

The number of topology-changing updates increases with order, introducing an additional source of autocorrelations. This overhead is absent in grouped sampling, since it eliminates the need to discover diagrammatic topologies stochastically. Thus, by following our iterative procedure and having all diagrams of a given order known from the outset, one not only has access to the full diagrammatic content of the perturbation series, but this knowledge can also significantly facilitate numerical calculations by reducing the variance of stochastic estimators.

In addition, it is worth noting that the grouped sampling yields a more uniform computational workflow, in which the same set of diagrammatic expressions is evaluated in parallel across different internal configurations. Such a structure favors modern computing architectures, for which applying identical operations to many data sets is more efficient than executing topology-dependent code paths, thereby providing an additional computational parallelization benefit without affecting the physical aspects of the final results.

\begin{figure}[t]
    \makebox[\textwidth][c]{%
        \includegraphics[width=18.5cm]{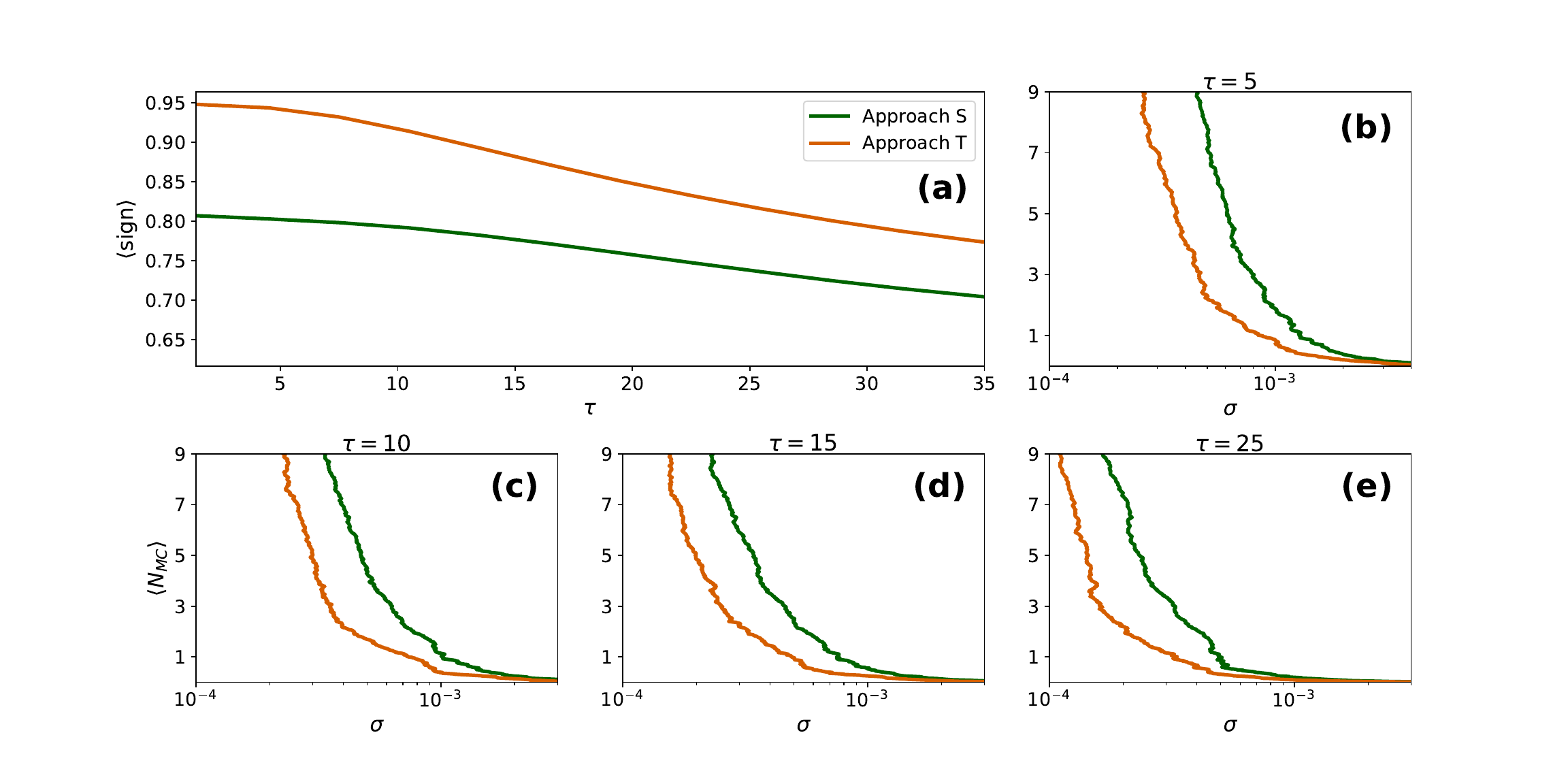}%
    }
    \caption{(a) Dependence of the average sign of the self-energy on the external imaginary time $\tau$ after $8.64\times{}10^{10}$ DMC updates; (b-e) Average number of DMC updates per hundred million needed to reach a given accuracy, i.e, a certain value of the standard deviation of the self-energy. The four panels correspond to 4 different external \nver{imaginary times $\tau = 5, 10, 15, 25$}. The curves are obtained by averaging over 96 independent DMC runs. The problem parameters used for calculations are $p = \pi$, $t_0 = 0.5$, $\mu = -0.8$, $\omega_0 = 0.8$, corresponding to the external (polaron) momentum, the electron hopping parameter, the chemical potential, and the Debye energy, respectively.}
    \label{fig:myfigure}
\end{figure}

To illustrate the advantages of the grouped sampling, we consider a model with an interaction vertex that changes sign depending on the incoming and outgoing momenta, namely the BLF-SSH model, when a single electron is coupled to an acoustic phonon branch. The coupling is given by $g(k, q) \propto \sin(k+q) - \sin(k) $, with $k$ and $q$ denoting the incoming electron and outgoing phonon momenta, respectively. Although the sign problem for the description of the BLF-SSH polaron formation is not as severe as in many-fermion systems, this model provides a transparent setting to compare two diagram-sampling strategies: approach $S$, in which diagrams are sampled separately, and approach $T$, in which all diagrams of a given order are sampled together.

Specifically, we compare here the $S$ and $T$ Diagrammatic Monte Carlo calculations of the sixth-order contributions to the BLF-SSH polaron self-energy, involving 10 diagrams, assuming a one-dimensional lattice. The electron and the phonon imaginary-time Green functions at zero temperature are $G_k(\tau)=e^{-\varepsilon_k\cdot{}\tau}$ and $D_q(\tau)=e^{-\omega_q\cdot{}\tau}$, respectively, with the electron and the acoustical phonon dispersion being $\varepsilon(k) = t_0 (1-\cos(k))\allowbreak + \mu$  and $\omega(q) = \omega_0 \sin(|q|/2)$. The parameters for our calculations are specified in the caption of Fig.~\ref{fig:myfigure}.

For the BLF-SSH polaron problem, the dependence of the average sign on the external imaginary time $\tau$ is shown in Fig.~\ref{fig:myfigure}(a). The average value is based on 96 independent DMC runs with $9\times{}10^{8}$ DMC updates. While the sign problem is somewhat enhanced at larger values of $\tau$, it remains mild, as expected. Nevertheless, even under these conditions, one can clearly observe that grouped sampling further suppresses sign fluctuations, with the $T$ curve remaining closer to the value $1$, which corresponds to the absence of the sign problem.

The minimal average number of DMC updates needed to achieve a given level of accuracy, characterized by the standard deviation of the self-energy, is shown in Figs.~\ref{fig:myfigure}(b-e). The advantages of grouped sampling are clearly evident in comparing the $T$ and $S$ curves. The latter reaches the same values as the former only after significantly more DMC updates. For example, in Fig.~\ref{fig:myfigure}(e), a standard deviation of $\sigma = 2\times 10^{-4}$ is reached by the $S$ curve after approximately a factor of four more DMC updates than in the $T$ case. This factor increases further when higher accuracy is required. Moreover, this trend is expected to become even more pronounced in higher-order calculations, because the number of diagrams{\nver{blue}, which is the bottleneck of the S approach, grows with order faster than factorial}, as reported in Table~\ref{tab1}.}

\nver{It may be argued that the advantage of the $T$ calculations over the $S$ ones can be readily seen for large expansion order $n$ when the number of self-energy diagrams is $\mathcal{I}^{(n)}_{irr}$>(n/2)!, each containing $1+3n/2$ internal parameters. Single DMC updating of all internal parameters takes $\mathcal{I}^{(n)}_{irr}\times(1+3n/2)$ evaluations of single diagram both for $S$ and $T$ approaches because one has to update all $\mathcal{I}^{(n)}_{irr}$ diagrams for $S$ approach and calculate the same number of terms in the sum for $T$ approach. The advantage lies in the evaluation of the relative contributions of different diagrams of order $n$, because the relative weight of separate diagrams is already exactly accounted for in the $T$- sum approach. However, even a single mutual reweighting of the separate diagrams takes at least $2\cdot{}(\mathcal{I}^{(n)}_{irr}-1)^2>((n/2)!)^2$ evaluations of a single diagram even for the one update to jump between different topologies, whereas much more than one such reweighting is necessary to reach the accurate result.} 

\section{Conclusions}
\label{sec7}

We have reduced the problem of polaron formation to determining the exact vertex function and deriving an expression for the electron self-energy in terms of a nested continued fraction involving the bare electron propagator and the exact vertex function. Building on this insight, we developed a method to iteratively construct all self-energy diagrams by combining noncrossing and vertex-correction diagrams from previous lower-order steps. Starting from the leading-order self-energy diagram, our simple four-step procedure (outlined in Section~\ref{SecSS}) systematically and straightforwardly generates higher-order diagrams. As a consistency check, we derived expressions to count the number of diagrammatic contributions to the exact vertex function, self-energy, and electron propagator at any order in the expansion.

\nver{To establish the completeness of our method, we introduce Flajolet's combinatorial formalism of Stieltjes-type continued fractions into Feynman diagrammatics. The SCBA self-energy is written as a continued fraction and expanded into Stieltjes-Rogers polynomials, whose combinatorial structure is encoded by Dyck paths. This yields a bijection between Dyck paths and SCBA diagrams, with each monomial term corresponding to a unique diagrammatic contribution. The construction extends naturally to vertex corrections, leading to an expansion in one-to-one correspondence with diagrams containing vertex insertions. Together with the Ward-Takahashi identity, this provides a diagrammatic rule that generates the complete set of diagrams at arbitrary order. We further identify two elements of diagram construction that allow for generalization to finite-electron-density cases.

Our method provides a particularly efficient way to navigate the complexity of high-order diagrammatic expansions, offering both theoretical insights and practical utility. In particular, it inherently accounts for all self-energy diagrams in a given order, thereby considerably improving the convergence of numerical techniques by reducing the most time-consuming computational effort required to determine the correct relative contributions of diagrams of different topologies. Thus, similar combinatorial techniques may be applied to a broader range of many-body problems, including those with finite electron densities and the renormalization of boson/interaction propagators.}

\section*{Acknowledgements}
\paragraph{Funding information}
\nver{J.K. and O.S.B. acknowledge the support of the Croatian Science Foundation under the project numbers IP-2022-10-3382 and IP-2022-10-9423, respectively. J.K. also acknowledges the support of the Croatian Science Foundation under the project number UIP-2025-02-5952. A.S.M. and S.R. acknowledge the support of the Croatian Science Foundation under the project number IP-2024-05-2406.} T.M. acknowledges the support of Project FrustKor financed by the EU through the National Recovery and Resilience Plan 2021-2026 (NRPP). \nver{O.S.B.
acknowledges the support of the project “Implementation
of cutting-edge research and its application as part of the
Scientific Center of Excellence for Quantum and Complex
Systems, and Representations of Lie Algebras,” Grant No.
PK.1.1.10.0004, co-financed by the European Union through the European Regional Development Fund-Competitiveness
and Cohesion Programme 2021-2027.}

\begin{appendix}

\section{Ward-Takahashi identity in terms of product of poles\label{App1}}

We prove by induction that Eq.~(\ref{WTid}) holds for a product of any number of poles. First, we consider the base case $l = 1$, which can be shown to hold using simple algebra,

\begin{equation}
\frac{1}{\omega-\omega_1-\omega'}-\frac{1}{\omega-\omega_1} = \frac{\omega'}{(\omega-\omega_1)(\omega-\omega_1-\omega')}\;.
\label{A1}
\end{equation}

\noindent Assuming that Eq.~(\ref{WTid}) holds for some $l\in{}\mathbb{N}$, it is sufficient to show now the same property for $l\to{}l+1$. Starting with the RHS of Eq.~(\ref{WTid}), we derive the LHS of Eq.~(\ref{WTid}) for the case when the number of poles equals $l+1$. Decomposing the  summation for this case into a sum from 1 to $l$ while treating the $(l+1)$-th contribution separately, one gets

\begin{equation}
\begin{aligned}
&\sum_{i=1}^{l+1}\,\frac{\omega'}{(\prod_{n<i}(\omega-\omega_n))(\omega-\omega_i)(\omega-\omega_i-\omega')\prod_{n'>i}(\omega-\omega_{n'}-\omega')} = \frac{1}{\omega-\omega_{l+1}-\omega'}\cdot{}\\
&\quad{}\quad{}\quad{}\cdot{}\sum_{i=1}^{l}\,\frac{\omega'}{(\prod_{n<i}(\omega-\omega_n))(\omega-\omega_i)(\omega-\omega_i-\omega')\prod_{n'\in{}\langle{}i, \left.l\right]_{\mathbb{N}}}(\omega-\omega_{n'}-\omega')}+\\
&\quad{}\quad{}\quad{}\quad{}\quad{}\quad{}\quad{}+\frac{\omega'}{(\prod_{i = 1}^{l}(\omega-\omega_{i}))(\omega-\omega_{l+1})(\omega-\omega_{l+1}-\omega')}\quad{}.
\label{A6}
\end{aligned}
\end{equation}

\noindent Assuming that Eq.~(\ref{WTid}) holds for $l$,  the RHS of Eq.~(\ref{A6}) may be rewritten in the form given by

\begin{equation}
\begin{aligned}
=&\frac{1}{\omega-\omega_{l+1}-\omega'}\cdot{}\left(\frac{1}{\prod_{i=1}^{l}(\omega-\omega_i-\omega')}-\frac{1}{\prod_{i=1}^{l}(\omega-\omega_i)}\right)+\\&\quad{}\quad{}\quad{}+\frac{\omega'}{(\prod_{i = 1}^{l}(\omega-\omega_i))(\omega-\omega_{l+1})(\omega-\omega_{l+1}-\omega')}\;,
\end{aligned}
\label{A7}
\end{equation}

\noindent which straightforwardly gives Eq.~(\ref{WTid}) for $l+1$, completing the proof.
\end{appendix}





%

\bibliography{bibliography}

\end{document}